\documentclass[nonacm]{acmart}
\AtBeginDocument{%
  }

\usepackage{booktabs}
\usepackage{makecell}   
\usepackage{footmisc}   
\usepackage{threeparttable}
\usepackage{url}
\usepackage{hyperref}

\usepackage{array}
\newcolumntype{L}[1]{>{\raggedright\arraybackslash}p{#1}}
\newcolumntype{C}[1]{>{\centering\arraybackslash}p{#1}}

\begin{document}

\title{From TCP/IP to MCP: Security and Governance Lessons from Internet Protocols for Agentic AI}

\author{Ben Bucknall}
\email{bucknall@robots.ox.ac.uk}
\affiliation{%
  \institution{University of Oxford}
  \country{UK}
}

\author{Andrew Byrley}
\affiliation{%
  \institution{Independent}
  \country{USA}
  }

\author{Lewis Hammond}
\affiliation{%
  \institution{Cooperative AI Foundation}
  \country{UK}
}

\renewcommand{\shortauthors}{Bucknall et al.}

\begin{abstract}
The past two years have seen considerable interest in protocols for managing how AI agents interact with human users, external tools, and each other.
In many ways, these protocols, including the \emph{Model Context Protocol} (MCP) and \emph{Agent2Agent} (A2A), mirror the suite of protocols that underlie the modern internet.
Motivated by this parallel, this paper draws lessons from the development, history, and governance of internet protocols with the aim of informing the responsible development of AI agent protocols.
Based on our analysis, we make two conclusions. First, if AI agent protocols do not sufficiently prioritise security in the early stages of development, they risk accumulating a security deficit that becomes increasingly costly to address over time due to practical and economic challenges.
Second, current governance structures for AI agent protocols centralise decision-making power in a small number of individuals and companies, raising important questions regarding legitimacy, especially if they become critical digital infrastructure.
We conclude with a discussion and implications for policymakers and developers of AI agent protocols.
\end{abstract}

\maketitle

\section{Introduction}
\label{sec:Intro}
AI agent protocols are emerging rapidly as a way to enable interoperability between agents, external tools and users. A protocol is a standardised set of rules governing the format and exchange of data between systems; in practice, the boundary between a `protocol' and a `standard' for technology is blurred, with protocols often becoming standardised through adoption. The predominant example, the Model Context Protocol (MCP), released by Anthropic in November 2024, has grown rapidly with over 170,000 supported tools \citep{stein_how_2026} and backers including OpenAI, Amazon, and Google. MCP is increasingly referred to as being in the process of becoming standardised digital infrastructure \citep[see e.g.][]{nisha_how_2025}.

These protocols functionally mirror those developed at the end of the 20th century that underlie the internet, with both facilitating communication and interoperability between heterogeneous systems. Inspired by this parallel, this paper draws lessons from internet protocols to inform the development of AI agent protocols. We focus on security and governance of protocols as areas where internet and agent protocols share structural similarities despite content-level differences. We explicitly exclude considerations of geopolitics in standard-setting, while acknowledging that standards and protocols development can be an active arena for advancing geopolitical power \citep{denardis_protocol_2009, faaborg-anderson_geopolitics_2022}.

While the comparison can be instructive, it is not perfect. In particular, internet protocols emerged from publicly-funded research over decades, while AI agent protocols have been developed commercially in months. However, the rapid rate of uptake of AI agent protocols only increases the timeliness of any lessons that can be learnt from the history of the internet. As these protocols become embedded in how AI systems interact with the world, decisions made now about their design, security, and governance will shape who benefits, who bears the risks, and who gets a say.

Throughout the paper, we draw on MCP as a running example due to its current position as the most widely-used and most-developed AI agent protocol. While other AI agent protocols are available, such as Agent2Agent and the Agent Communication Protocol, MCP offers the greatest evidence base from which to draw a comparison to internet protocols. When considering protocol governance, we primarily focus on the Internet Engineering Task Force (IETF) given its stewardship of the majority of internet protocols, including TCP, IP, HTTP, TLS/SSL, and DNS.

Based on our analysis, we make two primary conclusions:
\begin{itemize}
    \item Retrofitting security measures to existing protocols is challenging due to practical and economic challenges. Early internet protocols deprioritised security, contributing to a decades-long security debt. If AI agent protocols do not sufficiently prioritise security in the early stages of their development, they risk accumulating a similar security deficit that becomes increasingly costly to address as adoption grows.
    \item Internet protocols are embedded in a unique governance structure with formal mechanisms for accountability and community input, including at the highest levels of leadership, though whether these mechanisms translate to openness in practice is disputed. In contrast, AI agent protocols have emerged with a governance structure that centralises decision-making power in a small number of individuals, raising concerns about legitimacy in the case that these protocols become critical digital infrastructure.
\end{itemize}

Alongside sections motivating each of these conclusions (Sections \ref{sec:Security} and  \ref{sec:Governance}, respectively), Section \ref{sec:Background} provides relevant background information and Section \ref{sec:DiscussionConclusion} discusses the two conclusions, how they interact, and highlights concrete implications for key stakeholders.

\section{Background}
\label{sec:Background}

\subsection{A Short History of Internet Protocols}
\label{sec:History}
The internet's origins lie in the Advanced Research Projects Agency Network (ARPANET), a U.S. Department of Defense project first operational in late 1969, designed to enable resource-sharing between geographically dispersed research computers \cite{leiner_brief_2009}. Critically, the host-to-host communication protocol was designed not by ARPA officials or by the contractors that built the network hardware, but by an informal group of graduate students who called themselves the Network Working Group (NWG) \citep{pelkey_interview_1988}. The NWG designed the Network Control Program (NCP), ARPANET's first host-to-host protocol \citep{crocker_official_1970}, which assumed a single, homogeneous network.

NCP was unsuitable for internetworking. In the early 1970s, DARPA was building packet radio and packet satellite networks that NCP could not accommodate due to differing packet sizes, latencies, and failure modes \citep{leiner_brief_2009, cerf_specification_1974}. This problem was solved in September 1981 by a pair of protocols: the Internet Protocol (IP) \citep{postel_internet_1981}, a connectionless datagram protocol responsible for addressing and routing, and Transmission Control Protocol (TCP) \citep{postel_transmission_1981}, a reliable byte-stream transport. On January 1, 1983 NCP was permanently disabled on ARPANET and TCP/IP became the sole operational protocol \citep{leiner_brief_2009}. By the time that commercial restrictions on internet traffic were lifted in 1995, TCP/IP was already the foundation of global networking \citep{leiner_brief_2009}, catalysed by the introduction of the World Wide Web in August 1991 \citep{berners-lee_hypertext_1996}.

The institutional home for internet protocol development, the Internet Engineering Task Force (IETF), grew directly out of this ARPANET community and was spun off as an engineering-focused task force in January 1986. The IETF's cultural charter was reflected by David Clark at the 1992 Cambridge plenary: `We reject: kings, presidents and voting. We believe in: rough consensus and running code' \citep{clark_cloudy_1992}. Unlike traditional standards-setting organisations, governments play no formal role in the IETF, and as such, it lacks national delegations, treaty obligations, and reserved government seats. We return to the details of the IETF's governance structure, and its contrast with MCP's, in Section \ref{sec:IETF}.

The primary vehicle for IETF standards development is the Request for Comments (RFC) series, which predates the IETF itself. The name `Request for Comments' was deliberately chosen to signal informality, reflecting `the basic ground rules [...] that anyone could say anything and that nothing was official' \citep{crocker_origins_1987}. What began as informal notes among a handful of ARPANET researchers became the authoritative document series for internet standards. Along with technical specifications, RFCs also encompass governance and process documents,\footnote{For example, RFC 2026 \citep{bradner_internet_1996}, defining the standards process itself, and RFC 8713 \citep{kucherawy_iab_2020}, governing leadership selection.} philosophical statements,\footnote{RFC 3271, `The Internet is for Everyone' \citep{cerf_internet_2002}; RFC 1958, `Architectural Principles of the Internet' \citep{carpenter_architectural_1996}.} ethical guidelines,\footnote{RFC 1087, `Ethics and the Internet' \citep{iab_ethics_1989}.} humour,\footnote{In particular, the April Fools' tradition, dating to 1978, includes RFC 1149, `A Standard for the Transmission of IP Datagrams on Avian Carriers' \citep{waitzman_standard_1990}.} and the RFC process itself.\footnote{RFC 2223, `Instructions to RFC Authors' \citep{postel_instructions_1997}.}

The principal competitor to TCP/IP in the 1980s was the `Open Systems Interconnection' (OSI) model, developed through formal international standards processes. OSI was developed by the International Organization for Standardization (ISO) and the International Telegraph and Telephone Consultative Committee (CCITT, now the ITU-T -- the Telecommunication Standardisation Sector of the International Telecommunication Union). It enjoyed powerful institutional backing, being endorsed by the European Community, national post, telephone, and telegraph administrations (PTTs) \citep{russell_rough_2006}, and major vendors including IBM and DEC \citep{russell_internet_2013}. Yet TCP/IP was already deeply entrenched in the research and university communities and its user base was growing rapidly \citep{leiner_brief_2009}.

The two efforts embodied different approaches to protocol development. The IETF's process required that protocol proposals be accompanied by working implementations; specifications evolved through iterative testing and deployment, not committee negotiation. OSI, by contrast, aimed to produce complete, formally specified standards through a multi-year, delegation-based process involving national standards bodies before implementations existed. The consequences of these competing approaches were significant. The need to reconcile competing interests within ISO produced enormous specification complexity, which made interoperable implementation difficult in practice \citep{russell_rough_2006}. By the early 1990s, many of the ISO committees working on OSI standards were coordinating their own work over TCP/IP-based email -- conducting their deliberations on the very infrastructure whose intended replacement they were designing \citep{russell_rough_2006}. Amplified by the explosive growth of the World Wide Web after 1993, displacement of TCP/IP became effectively impossible \citep{russell_rough_2006} and mandates for the adherence to the OSI model faded in both the US and EU. OSI as a deployed protocol suite effectively ceased to exist \citep{russell_rough_2006}, although individual OSI-derived standards survived in specific niches.\footnote{Most notably, X.509 digital certificates, which remain foundational to internet security.}

\subsection{AI Agent Protocols to Date}
\label{sec:AIAProtocols}
Similarly to how internet protocols are specifications for mediating connections between individual users and machines, AI agent protocols specify the structure of how AI agents interact with external tools, human users, and each other. By standardising these interactions, protocols serve to make it easier for individual agents to reliably perform actions in the real world, while also providing the opportunity for attribution, oversight, and certification \citep{chan_infrastructure_2025}. While protocol development for AI agents has progressed rapidly in recent years, the concept was discussed much earlier, for example in \citet{wooldridge_introduction_2009}, who characterises protocols as defining `the ``rules of encounter'' between agents.'

Numerous efforts to develop AI agent protocols are underway within government, industry, and academia. For example, the National Institute for Standards and Technology (NIST) in the US has launched an AI Agent Standards Initiative to promote the trusted adoption of AI agents and `[cement] U.S. dominance at the technological frontier'.\footnote{\url{https://www.nist.gov/news-events/news/2026/02/announcing-ai-agent-standards-initiative-interoperable-and-secure}. Accessed 6/10/2026.} So far, the initiative has supported a concept paper on `Software and AI Agent Identity and Authorization' \citep{booth_accelerating_2026}, and a request for comments regarding `Security Considerations for AI Agents'.\footnote{\url{https://www.federalregister.gov/documents/2026/01/08/2026-00206/request-for-information-regarding-security-considerations-for-artificial-intelligence-agents}. Accessed 6/10/2026.} Elsewhere, the ITU has started developing standards specifically focussed on AI agents \citep{itu_requirements_2025}, as has the China Academy of Information and Communications Technology \citep{van_draanen_glismann_lexicon_2025}. Finally, while its standards to-date are not specifically focused on AI agents, much of the ISO's work on AI -- for example throught its ISO/IEC JTC 1/SC42 committee -- is directly relevant \citep{nannini_ai_2026}.

A larger number of proposed protocols have come from industry and academic sources, variously focussing on agent-agent, agent-user, or agent-tool interactions \citep{yang_survey_2025}. While such proposals don't have the explicit aim of defining a formal standard, as seen in the case of internet protocols, it is not uncommon for informal proposals to become \textit{de facto} standardised if they secure enough buy-in among users and developers. Examples include Agent2Agent (A2A) and Agent Communication Protocol (ACP),  released by Google in April 2025 and IBM in March 2025, respectively, both of which focus on inter-agent communication and collaboration \citep{surapaneni_announcing_2025, martineau_open-source_2025}. Others, such as AG UI, developed by CopilotKit focuses on the interaction between user interfaces and AI agent-based backends.\footnote{\url{https://docs.ag-ui.com/introduction}. Accessed 6/10/2026.} While many of these examples are stewarded and supported by private industry actors, they take the form of open-source projects with a focus on developing functional implementations of protocols rather than abstract frameworks.

The Model Context Protocol (MCP) has emerged as the predominant AI agent protocol since its release by Anthropic in November 2024. Framed as a `USB-C port for AI applications', MCP provides a standardised format for AI agents to interact with external tools as a way of addressing the previously fragmented landscape, where each pair of agent and tool might have a unique way of interacting \citep{yang_survey_2025, ehtesham_survey_2025}. Following its release, MCP quickly gained traction within the community and corporate buy-in, notably with OpenAI supporting MCP across its products from March 2025.\footnote{\url{https://techcrunch.com/2025/03/26/openai-adopts-rival-anthropics-standard-for-connecting-ai-models-to-data/}. Accessed 6/10/2026.} The MCP ecosystem has grown to include tens of thousands of servers -- lightweight programs that expose tools to agents -- and hundreds of thousands of distinct tools \citep{stein_how_2026}. In December 2025, Anthropic donated MCP to the Agentic AI Foundation (AAIF), a subsidiary of the Linux Foundation, which it founded along with other companies including OpenAI and Google. We discuss the AAIF and its governance structure and implications further in Section \ref{sec:Governance}.

\subsection{On the Comparison of Internet and AI Agent Protocols}
\label{sec:Comparison}
Having provided an overview of the history of internet protocols and ongoing developments in AI agent protocols, we now discuss in more detail the motivation for making this comparison. We outline similarities regarding what purpose the protocols are intended to serve, how they're developed, and by whom, as well as ways in which they contrast.

The similarity between internet protocols and AI agent protocols at a functional level is that both aim to address problems of interoperability and heterogeneity within their domains. In particular, TCP/IP standardised the format in which data could be transferred between computers of varying architectures, running different operating systems. In comparison, AI agent protocols standardise the format in which agents interact with users, tools, or other agents. For example, rather than AI agents needing to structure a tool call in a different way for each tool, MCP defines a single format that can be used with a large number of potential tools. As a result of this focus on interoperability and interconnectivity, both kinds of protocols are subject to network effects whereby the benefits of an individual adopting the protocol are proportional to the number of existing adopters \citep{kuerbis_hidden_2020, ozment_bootstrapping_2006}.

Furthermore, in both the case of TCP/IP and industry-led AI agent protocols, the specification is approved by a centralised authority, though amendments, updates, or extensions can be proposed by the broader community. The approved standard then serves as reference for implementation by third-parties that are nonetheless able to deviate from the specification. This means that the success of standardised protocols stems from soft power and network effects rather than any mandated use, and that the individual or group with power to define the specification exercises significant power over the ecosystem without any formalised authority.

At an ecosystem level, both internet and AI agent protocols are advanced by a range of organisations and institutions. As taxonomised by \citet{biddle_no_2016}, these include `formal standards development organisations' (SDOs) such as the ISO and ITU and `consortia' -- industry-led organisations that lack government recognition or authority -- such as the IETF and AAIF. \citet{biddle_no_2016} also discusses `single promoter standards' -- standards that are developed and promoted by a single company. Anthropic's initial development of MCP can be seen as an example of this category.

Alongside the similarities, there are also numerous differences worth noting. First, while early internet development was overwhelmingly supported and funded by the US Department of Defense through DARPA, (implemented) AI agent protocols are almost entirely the result of private, corporate funding. As such, while the internet was initially intended as an inherently national security-relevant technology, rather than one used by the broad public, companies developing AI agent protocols may view them as a commercial asset, product, or service. Indeed, MCP, the first widely-adopted AI agent protocol, was initially developed by a single private company.

Second, AI agent protocols have experienced a dramatically faster rate of uptake than the internet. \citet{stein_how_2026} identifies over 70,000 potential MCP servers as of February 2026 -- approximately 15 months after MCP's initial launch. In comparison, it took approximately two decades for the internet to reach a comparable number of hosts \citep{zakon_hobbes_1997}. This may be a direct result of the difference in initial purpose noted above, in that the internet was initially a closed community of academics developing defence infrastructure. While the user base of AI agent protocols remains highly technically-literate, there is uncertainty regarding whether this will shift to encompass a broader, less-technical subset of the general public, as was seen for the internet.

Overall, while these differences mean that the analogy is imperfect, the structural similarities between internet and AI agent protocols -- both at the functional and ecosystem level -- suggest that the comparison is informative. In the following two sections we compare the development of internet and AI agent protocols, particularly with regards to their security properties and governance structures.

\section{The Challenges of Retrofitting Security}
\label{sec:Security}
In this section we discuss the history of internet security, and how excluding security as an initial design goal created challenges when it came to retroactively implementing security measures -- that is, technical defences against the misuse of the internet for malicious ends. We also provide a brief overview of security in MCP, arguing that, if neglected in these early stages of MCP's development, implementing security will face similar challenges. As a way of structuring the discussion, we introduce two complementary categories of \textit{practical} and \textit{economic} challenges to retrofitting security.

\textit{Practical} challenges occur when, despite the existence of secure protocols, implementation is bottlenecked by material constraints. This may be due to implementation costs or performance trade-offs. For example, performing cryptographic verification can introduce a small but real computational overhead to communication protocols \citep{naylor_cost_2014}. Having to retroactively implement security to existing protocol suites raises additional practical challenges, such as minimising service disruption, as well as ensuring interoperability and backwards-compatibility \citep{barker_considerations_2025}. Moreover, maintaining backwards compatibility with insecure legacy systems can itself introduce new attack vectors that exploit continued support for deprecated protocol versions \citep{jager_one_2013}.

\textit{Economic} challenges occur when implementing security, while not theoretically or practically challenging, is disincentivised or inhibited. For example, uptake of secure protocols may be slowed by collective action problems whereby, despite manageable implementation costs, the benefit to each actor is negligible unless all other actors also implement the secure protocol \citep{kunreuther_interdependent_2003, bauer_cybersecurity_2009, camp_system_2004, gordon_externalities_2015}. \citet{anderson_economics_2006} provide an influential overview of the economic factors at play in information security, arguing that `incentives are becoming as important as technical design in achieving dependability.' These challenges are compounded when security must be retrofitted, as secure protocols must then compete with insecure alternatives that are already broadly deployed. For example, actors may anchor on the cost of an insecure `status quo', making them less likely to pay extra for security, even if they would have paid the higher price if no alternative was available \citep{samuelson_status_1988, kim_investigating_2009, polites_shackled_2012, redmiles_dancing_2018}.

\subsection{The Internet's Persistent Security Deficit}
\label{sec:InternetSec}
As discussed in the previous section, the modern internet emerged out of a small and closed network of trusted users, predominantly in academia, in the 1970s \citep{denardis_history_2007}. At the time, personal computers were not yet commonplace, and `public or home access [to the internet] seemed implausible' \citep[][p. 683]{denardis_history_2007}. As such, security was not considered to be a priority. On the contrary, it was deemed to be both infeasible and undesirable \citep{eisenberg_cornell_1989}:
\begin{quote}
A community of scholars should not have to build walls as high as the sky to protect a reasonable expectation of privacy, particularly when such walls will equally impede the free flow of information. Besides, attempting to build such walls is likely to be futile in a community of individuals possessed of all the knowledge and skills required to scale the highest barriers.
\end{quote}

A watershed moment came in November 1988 when the then-Cornell graduate student Robert Morris developed and released a self-replicating computer program (what's now termed a `worm') on the internet, disabling over 6,000 computers\footnote{Thought to be approximately 10\% of all internet-connected computers at the time \citep{denardis_history_2007}.} \citep{eisenberg_cornell_1989, denardis_history_2007}. The `Morris Worm', as it has since become known, exposed the weaknesses in relying on interpersonal trust and academic integrity for securing a growing network. Around the same time, Steve Bellovin published a paper outlining multiple `serious security flaws inherent in the [TCP/IP suite of] protocols' \citep{bellovin_security_1989}. For example, since TCP/IP lacked a method for verifying the stated source address of packets sent over the network, attackers could impersonate a trusted host and, in combination with predictable protocol behaviour, execute commands on remote systems without authentication \citep{bellovin_security_1989}. Over the ensuing decade, internet user numbers rose rapidly, catalysed in part due to the advent of the World Wide Web in the early 90s, with many new users coming from outside the academic communities which comprised much of the early user base. As a result, the heavily trust-based model on which the early internet operated became outdated. Combined with commercial and corporate interest in e-commerce, the need for greater internet security was becoming apparent \citep{denardis_history_2007}. Costly worms and viruses underscored this need, and remained commonplace into the 2000s, with examples such as the `Love Bug' virus (2000), `Code Red' (2001), and `Mydoom' (2004) requiring billions of dollars to combat \citep{denardis_history_2007}.

In spite of the rising costs of security vulnerabilities, adoption of more secure protocol variants has historically proven sluggish. One prominent example is that of retrofitting encryption to the Hypertext Transfer Protocol (HTTP) -- the protocol underlying the World Wide Web. HTTP/1.0 was published as a standard in 1996 as RFC 1945 \citep{berners-lee_hypertext_1996}, having gained broad adoption during its draft status. However, HTTP did not natively support encryption, meaning that all information transferred using it was liable to man-in-the-middle attacks such as eavesdropping, tampering, or forgery.

Around the same time, Netscape was developing the Secure Sockets Layer (SSL), a proprietary protocol for cryptographically securing the transmission of data over TCP connections, with HTTP being the primary motivation. SSL was later adopted by the IETF, rebranded to Transport Layer Security (TLS), and published in RFC 2246 in 1999 \citep{dierks_tls_1999}. `HTTP over TLS' or simply `HTTPS' was standardised in 2000 as RFC 2818 \citep{rescorla_http_2000}. Despite all technical pieces being in place by the late 1990s, HTTPS adoption remained confined to a narrow set of use cases -- primarily financial transactions, where payment card industry requirements mandated its use \citep{ozment_bootstrapping_2006}. The broader web remained unencrypted for two decades. In 2017, half of the most popular websites were still HTTP by default, and most smaller websites did not support HTTPS at all \citep{felt_measuring_2017}. The barriers to greater uptake identified in the literature were practical and economic rather than technical. For example, it was costly to acquire a TLS certificate -- a prerequisite for allowing users to connect using HTTPS -- and deployment was a `notoriously difficult process' \citep{bernhard_usability_2019} deemed to be `far too complex even for people with proficient knowledge in the field' \citep{krombholz_i_2017}. With unencrypted HTTP as the entrenched default and minimal direct costs to web servers for not providing secure connections, the additional burden of deploying HTTPS proved prohibitive for many smaller actors. It took structural changes such as \textit{Let's Encrypt} providing a cheaper and simpler method for acquiring TLS certificates \citep{aas_lets_2019}, browsers warning users when connecting to insecure websites \citep{felt_rethinking_2016}, and the US government requiring all federal websites to support HTTPS by the end of 2016 \citep{scott_policy_2015} to finally make the use of HTTPS widespread.

While many similar cases have been individually documented, when read together they exhibit a common pattern of widespread adoption of secure internet protocols being hindered by practical and economic challenges, long after the technical substrate was in place. IPv6, despite being standardised as a successor to IPv4 in 1998 \citep{deering_internet_1998}, still only serves around 50\% of internet users\footnote{\url{https://www.google.com/intl/en/ipv6/statistics.html}. Accessed 6/10/2026.} due to a lack of backwards compatibility imposing significant costs on early adopters \citep{kuerbis_hidden_2020, wu_transition_2013, denardis_protocol_2009}. Domain Name System Security Extensions (DNSSEC) -- the suite of protocols that adds cryptographic attestation to the directory that maps domain names to IP addresses -- was introduced in 2005 \citep{arends_dns_2005, arends_resource_2005, arends_protocol_2005}, yet remains at single-digit adoption rates,\footnote{\url{https://stats.dnssec-tools.org/\#/?top=tlds\&tld_tab=3}. Accessed 6/10/2026.} partly due to collective action problems \citep{ozment_bootstrapping_2006}. Finally, Resource Public Key Infrastructure (RPKI) \citep{lepinski_infrastructure_2012}, a solution to insecurities in the Border Gateway Protocol (BGP), has only recently attained 50\% coverage, with adoption slowed by similar organisational challenges \citep{gouda_ru-rpki-ready_2025}.

A common theme among all these examples (HTTPS, IPv6, DNSSEC, RPKI) is that, despite the technology being functional, practical and economic considerations -- such as coordination problems and large upfront costs to first-movers -- considerably slowed the uptake of security. Often this was further hindered by secure protocols having to compete with existing but insecure alternatives, which could increase switching costs, demand backwards compatibility, or serve as an anchor against which performance is measured. In many cases it took structural changes in the form of regulatory mandates \citep{scott_policy_2015}, free or automated tooling \citep{aas_lets_2019}, and changes to default behaviour by major platform providers to achieve widespread use of secure protocols \citep{felt_rethinking_2016}. As we argue in the following section, there is reason to believe that these dynamics will generalise to AI agent protocols.

\subsection{Security Challenges for AI Agent Protocols}
\label{sec:AISec}
Despite its relative nascency, MCP has already experienced a number of security issues and flaws, and a considerable amount of research has been carried out analysing its security profile \citep[see e.g.][]{hou_model_2026, radosevich_mcp_2025, errico_securing_2025, anbiaee_security_2026, guo_systematic_2025, zhao_when_2025, croce_trivial_2025}. For example, \citet{hou_model_2026} formalise a threat taxonomy according to four potential attacker types: malicious developers, external attackers, malicious users, and security flaws. Notably, they find that, `despite the rapid adoption of MCP, its ecosystem is still in the early stages, with key areas such as security, tool discovery, and remote deployment lacking comprehensive solutions' \citep{hou_model_2026}. Examples of the security issues highlighted include `command injection', where an attacker inserts malicious logic into a server's source code, and `unauthorized access' attacks where attackers are able to interact with systems beyond intended permissions due to `weaknesses in authentication, transport, or session management' \citep{hou_model_2026}. Instances of these vulnerabilities have been observed. For example, an unofficial Postmark email MCP server `built trust over 15 versions' before secretly adding a backdoor that 'BCC'd emails to an external server.'\footnote{\url{https://postmarkapp.com/blog/information-regarding-malicious-postmark-mcp-package}. Accessed 6/10/2026.}

While MCP acknowledges potential security issues, current approaches to dealing with risks to safety and privacy place the responsibility on downstream developers and end users. For example, the protocol specification states that `users must explicitly consent to and understand all data access and operations' and that `users must retain control over what data is shared and what actions are taken.'\footnote{\url{https://modelcontextprotocol.io/specification/2026-07-28\#security-and-trust-\&-safety}. Accessed 8/9/2026.} Furthermore, while the MCP registry -- an official and open catalog of publicly available MCP servers -- includes a `community-driven mechanism for moderation',\footnote{\url{https://blog.modelcontextprotocol.io/posts/2025-09-08-mcp-registry-preview/}. Accessed 6/10/2026.} it defers security opportunities such as `source code scanning' and `tool poisoning countermeasures' to later iterations of the registry.\footnote{ \url{https://github.com/modelcontextprotocol/registry/discussions/11}. Accessed 6/10/2026.}

If not addressed in the near-term while MCP is still scaling, implementing solutions to many of these security weaknesses may prove challenging due to the practical and economic effects discussed in the previous section.
There has already been a case in which practical considerations, in the form of backwards-compatibility, played a role in MCP's evolution. Authorisation through OAuth \citep{hardt_oauth_2012} was first introduced to MCP as an optional feature in the March 2025 specification as a way of handling authorisation for accessing restricted servers.\footnote{\url{https://modelcontextprotocol.io/specification/2025-03-26/basic/authorization}.} Initially, the specification relied on Dynamic Client Registration (DCR) \citep{richer_oauth_2015} as a way of allowing clients to register with servers at runtime \citep{parecki_lets_2025}. However, DCR was seen to place too much onus on developers of MCP servers who often didn't have the necessary expertise to implement it correctly \citep{delimarsky_whats_2025}. In a study of real-world remote MCP servers that implement authorisation, \citet{zhou_first_2026} found that 96.6\% of tested servers exhibited a flaw with DCR, such as allowing attackers to intercept authorisation codes.
As such, the November 2025 MCP specification moved towards an alternative approach -- Client ID Metadata Documents (CIMD) -- but retains support for DCR as a fallback for the purposes of backwards compatibility.\footnote{\url{https://modelcontextprotocol.io/specification/2026-07-28/basic/authorization/client-registration}.} As MCP continues to grow, similar dynamics are likely to recur, with the difficulty of migrating to improved approaches -- and the security costs of maintaining legacy ones -- only increasing with adoption.

Furthermore, in April 2026, OX Security disclosed that the use of stdio transport for launching MCP servers as local subprocesses passes the command given in a client's configuration directly to the operating system without validation. This can facilitate arbitrary code execution if an attacker is able to influence the configuration \citep{ox_security_mother_2026}. Since this logic is built into the official MCP SDKs, it is inherited by all servers that are built on them. One of the proposed protocol-level fixes was to implement manifest-only execution, meaning that a client would only launch verified servers rather than executing an arbitrary command string. However, Anthropic responded by stating that this is not a vulnerability in MCP, since its trust model assumes that client configurations are trusted, and so the responsibility is on the host application's developer to implement safeguards. This position was clarified in an update to MCP's security policy.\footnote{\url{https://github.com/modelcontextprotocol/modelcontextprotocol/commit/677d6250ccf6731618e0228847565fdfceed9082}} The proposed fix would have broken the launch path for every existing stdio server, requiring server authors to publish conforming manifests and client implementers to enforce them, with the cost of coordination rising as the ecosystem grows.

Despite this section's focus on practical and economic security challenges, it is worth noting that, for many issues in AI agent security, we do not yet have robust technical solutions. Many of the observed vulnerabilities of MCP discussed above stem from weaknesses in agents themselves in the face of `agent traps' \citep{franklin_ai_2026}. For example, tool poisoning relies on LLMs' susceptibility to jailbreaking. These are analogous to social engineering attacks such as phishing, which exploit the human user rather than the protocol, but which the protocol nonetheless enables. In the internet case, however, the attacker-defender dynamic is grounded in the relative stability of human psychology. For example, innate human susceptibility to scams tends not to vary much. The same cannot be said of AI agents, whose underlying models are frequently updated. This means that the attack surface the protocol must contend with is itself shifting on the timescale of months, but also that security advances could come through updates to agents themselves. Combined with the field's still-limited understanding of LLM robustness \citep{kumar_robustness_2025}, this makes the residual risk that persists after protocol-level security measures less predictable than its internet-era analogue. Authentication, authorisation, and verification remain essential for MCP, but they are unlikely to be sufficient.


\subsection{Discussion}
\label{sec:SecurityDiscussion}
Above, we have argued that AI agent protocols are likely to face similar practical and economic challenges to implementing security measures as seen in the case of internet protocols. This argument largely follows from the observation in Section \ref{sec:Comparison} that both internet and AI agent protocols are subject to network effects -- once users have adopted a particular version of a protocol, updating the entire ecosystem to a newer, more secure version naturally encounters a coordination problem. While it is possible that MCP has already reached this stage given the tens of thousands of servers already in use, we do not empirically establish any existing security deficit in MCP. Rather, we have argued that the conditions under which such a deficit could emerge are present, and suggest empirical studies of security in the MCP ecosystem as valuable future work.

If it is the case that network effects are slowing the adoption of security in MCP, what can be learnt from internet protocols? The example of HTTPS discussed above points towards three events that spurred adoption: easy and free tooling (Let's Encrypt), changes to default system behaviour (browsers warning when connecting via non-encrypted HTTP), and mandated adoption (US federal websites requiring encryption). This suggests potential analogous interventions that could incentivise the uptake of secure AI agent protocols. For example, protocol developers or third-parties could develop public tooling that makes scanning and certification of MCP servers free and automated. Official MCP software development kits (SDKs) could include default behaviour that alerts the user if their agent is connecting to a tool server that has not been verified or certified. Major enterprises or government agencies could require that critical servers are verified or certified. Ultimately though, these interventions are able to incentivise the uptake of secure protocols only once they have been technically defined and incorporated into the protocol specification -- something that, for many critical security features, is still lacking.

\section{Governance and Legitimacy in Protocol Development}
\label{sec:Governance}
In this section, we explore and compare the governance and decision-making structures adopted by the dominant protocol development bodies responsible for developing internet and AI agent protocols -- namely, the Internet Engineering Task Force (IETF) and Agentic AI Foundation (AAIF), respectively. We find notable differences in approach. While the IETF has evolved over decades to adopt participatory approaches to both low-level protocols development and high-level strategic planning, the AAIF and MCP have adopted structures that centralise decision-making authority in a small number of industry representatives, with few inroads for community participation at the highest level.

\subsection{Internet Protocols: The IETF}
\label{sec:IETF}
As introduced in Section \ref{sec:History}, the IETF serves as the predominant body in internet standards and protocols development, with protocols including TCP/IP, HTTP, and SSL/TLS \citep{harcourt_internal_2020}. Initially formed in 1986 as a working group for the expanding set of internet users, particularly at universities, the IETF quickly gained traction and became open to all interested individuals within two years of its foundation \citep{russell_internet_2014}. This focus on openness, along with technical merit and action through broad consensus, has since been codified through RFCs as formal structures and processes for the body \citep[see e.g.][]{bradner_internet_1996, kucherawy_iab_2020, salz_entities_2022, bradner_ietf_1998, alvestrand_mission_2004, bradner_intellectual_2017}.

The openness of the IETF is represented by the fact that `there is no membership in the IETF.'\footnote{\url{https://www.ietf.org/about/introduction/\#participants}. Accessed 6/10/2026.} Participation, for example by signing up to a working group, is open to any interested individual. While all participation is considered by the IETF to be on a voluntary and individual basis, tech companies can, and do, employ individuals to contribute full-time, with many key protocols being proposed by contributors from a handful of companies \citep[][p. 38]{harcourt_internal_2020}. Decisions are made by rough consensus -- often indicated by humming -- rather than formal voting procedures, with working group chairs responsible for deciding when consensus is reached and ensuring that objections are addressed \citep{resnick_consensus_2014}.

\begin{figure*}[t]
    \centering
    \includegraphics[width=0.9\textwidth]{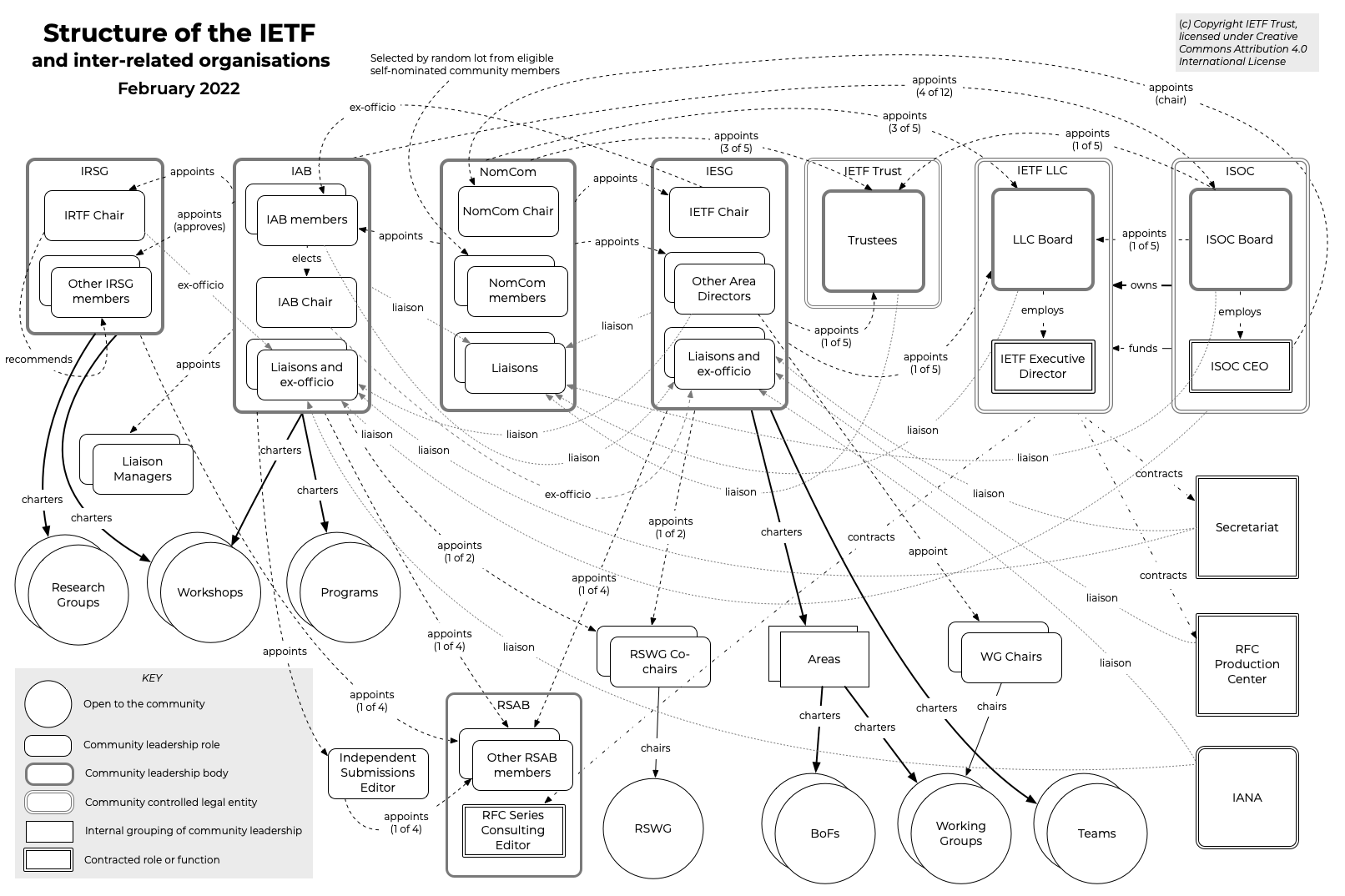}
    \caption{Structure of the IETF. By Jay Daley --
    \url{https://www.ietf.org/media/documents/IETF_Structure_-_February_2022.png}, CC BY 4.0, \url{https://commons.wikimedia.org/w/index.php?curid=144203249}}
    \label{fig:IETF}
\end{figure*}

While the technical ground work is carried out within working groups, there is a large amount of higher-level strategic and administrative work required to keep the IETF running (see Figure \ref{fig:IETF}). This is the purview of a number of bodies including: the Internet Architecture Board (IAB), the most senior technical body that sets long-term technical direction; the Internet Engineering Steering Group (IESG), which is responsible for the protocol development process, including providing final approval for RFCs; IETF Administration LLC (IETF LLC) which serves as the legal, financial, and operational centre; and the IETF Trust which manages the IETF's intellectual property \citep{salz_entities_2022}.

One of the most distinctive features of IETF's governance is the Nominating Committee (NomCom). Composed of 10 randomly selected\footnote{For example, as according to the process specified in \citet{eastlake_3rd_publicly_2004}.} contributors serving year-long terms, the NomCom has broad powers and responsibilities for appointing members to leadership positions across the IETF structure, including the boards of the IAB, IESG, and IETF LLC \citep{kucherawy_iab_2020}. Eligibility for the NomCom is subject to certain criteria, but can be satisfied, for example, by attending three of the past five IETF meetings \citep{duke_nominating_2023}. Furthermore, upon petition by 20 IETF members meeting the same eligibility criteria, a Recall Committee can be established according to the same processes as the NomCom with the power to recall (i.e. remove) any member of the IESG, IAB, or other NomCom-appointed leader by three-fourths majority \citep{kucherawy_iab_2020}.

However, despite adopting a formal governance structure that emphasises open participation, the IETF has received criticism, including for alleged corporate capture due to companies' employing individuals to participate in the standards setting process `as individuals' \citep{harcourt_internal_2020}. Furthermore, analysis has shown that formal openness may not translate to decentralisation, with 32\%-42\% of RFC drafts between 2010 and 2019 authored by the top 10\% most influential contributors \cite{khare_web_2022}. Critics have further pointed towards `cultural barriers to openness' in the IETF, including `confrontational working practices' that exclude input from non-profit organisations \citep{cath_send_2026}.

\subsection{AI Agent Protocols: MCP and the AAIF}
\label{sec:AAIF}
In this subsection, we provide the first comprehensive analysis of the governance structure surrounding MCP following this donation to the LF, drawing on charters, tier structures, and public governance documentation. We compare the structure that has emerged since MCP's founding donation to the \textit{Agentic AI Foundation} (AAIF) in December 2025 to other LF directed funds, namely the Cloud Native Computing Foundation (CNCF) and the PyTorch Foundation.

MCP itself follows a largely standard open-source project governance hierarchy consisting of: a pair of `lead maintainers', a `lifetime appointment' which holds `ultimate authority over MCP's direction and governance'; a group of six `core maintainers', holding `final decision-making authority for MCP's technical direction'; and a broader community of `maintainers', `members', and `contributors' (in descending order of responsibility) \citep{soria_parra_mcp_2026}. There is `no formal requirement' for becoming a contributor, other than contributing to MCP in any form \citep{soria_parra_mcp_2026}. While the positions of  `member' and `maintainer' are broadly attainable on a meritocratic basis, the core maintainer role is `intentionally limited' to ensure a `coherent technical vision while the project scales' \citep{soria_parra_mcp_2026}. Roughly in line with the `Benevolent Dictator For Life' (BDFL) model of open-source governance, the lead maintainer role is `a lifetime appointment reserved for project founders' which is only assumed through succession rules if an existing lead maintainer leaves the role \citep{soria_parra_governance_2025}.

The AAIF  is a `directed fund' under the Linux Foundation (LF) which stewards MCP alongside five further projects: OpenAI's `AGENTS.md' and Block's `goose',  which were donated at launch, as well as `agentgateway', `Agent2Agent' (A2A), and `Agent Router' which joined in June, August, and September 2026, respectively.\footnote{\url{https://agentgateway.dev/blog/2026-06-04-agentgateway-joins-aaif/}. Accessed 10/8/2026.\\ \url{https://aaif.io/blog/a2a-joins-aaif}. Accessed 8/9/2026.\\ \url{https://aaif.io/blog/agent-router-joins-aaif}. Accessed 6/10/2026.} The AAIF has a membership base of mostly private companies with some academic institutions, as is standard for LF directed funds. These members are organised according to tiers which provide differing levels of representation and input into high-level decision-making in exchange for increased membership fees. In the case of AAIF, these tiers are called `platinum', `gold', `silver', and `associate'.\footnote{This tiered structure is also common across LF directed funds. The PyTorch foundation has the same Platinum, Gold, Silver, and Associate tiers (\url{https://pytorch.org/foundation/}), whereas the CNCF has Platinum, Gold, and Silver, but splits the `Associate' category into separate `Academic', `Nonprofit', and `End User Supporter and Contributor' categories (\url{https://www.cncf.io/about/members/}). Both links accessed 6/10/2026.} The eight platinum members have the largest influence over the governance and direction of the AAIF and are currently: AWS, Anthropic, Block, Bloomberg, Cloudflare, Google, Microsoft, and OpenAI.\footnote{\url{https://aaif.io/members/}. Accessed 6/10/2026.} The AAIF charter specifies a `Platinum Member Cap' mandating that `at no time may there be more than eight (8) Platinum Members.'\footnote{\url{https://github.com/aaif/foundation/blob/main/foundation-charter.pdf}. Accessed 6/10/2026.} The current eight platinum members, as well as the platinum cap, have been in place since the launch of the AAIF.\footnote{See \url{https://web.archive.org/web/20251209184925/https://aaif.io/members/}.}

The AAIF is managed by a Governing Board (GB) holding responsibility over the budget, overseeing outreach, and reviewing proposals for additional projects to the foundation.\footnote{\url{https://github.com/aaif/foundation/blob/main/foundation-charter.pdf}. Accessed 6/10/2026.} Each platinum member is entitled to appoint a single representative to the GB, and gold members are collectively represented by a single representative. Silver and associate members have no representation on the GB -- a notable departure from other LF directed funds, including the PyTorch Foundation and the CNCF (see Table \ref{tab:gb-representation}).\footnote{The AAIF's community representation when compared to similar foundations contrasts with the GB chair's claim that the election of a single gold member representative to the GB strengthens `the foundation's commitment to inclusive, community-driven governance.' \citep{nalley_aaifs_2026}.} Alongside the GB, the AAIF establishes a Technical Committee (TC) for the purpose of overseeing best practices and the technical vision of the foundation.\footnote{\url{https://github.com/aaif/technical-committee/blob/main/README.md}. Accessed 6/10/2026.} As for the GB, each platinum member of the AAIF is entitled to appoint one representative to the TC.\footnote{\url{https://github.com/aaif/technical-committee/blob/main/governance/charter.md}. Accessed 6/10/2026.} There is no representation for gold, silver, or associate members on the TC, which again departs from other LF directed funds. For example, gold members of the PyTorch foundation have one representative on the `Technical Advisory Council' per three members, which are selected by the gold members themselves (see Table \ref{tab:tc-representation}).

\begin{table*}[t]
\centering
\caption{Governing Board representation per membership tier in the AAIF, CNCF, and
PyTorch Foundation. Sources: AAIF Charter, CNCF Charter, PyTorch Charter.}
\label{tab:gb-representation}
\begin{tabular}{L{1.8cm} L{2.8cm} L{3.8cm} L{3.8cm}}
\toprule
\multicolumn{1}{C{1.8cm}}{\textbf{Membership Tier}} &
\multicolumn{1}{C{2.7cm}}{\textbf{AAIF}} &
\multicolumn{1}{C{4cm}}{\textbf{CNCF}} &
\multicolumn{1}{C{4cm}}{\textbf{PyTorch}} \\
\midrule
\textbf{Platinum} &
1 representative from each member &
1 representative from each member &
1 representative from each member \\
\addlinespace
\textbf{Gold} &
1 representative &
1 representative per 5 members, up to a maximum of 3 &
1 representative per 3 members \\
\addlinespace
\textbf{Silver} &
None &
1 representative per 10 members, up to a maximum of 3 &
1 representative per 10 members, up to a maximum of 3 \\
\addlinespace
\textbf{Other} &
None &
The chair of the Technical Oversight Committee;\newline
The chair of the End User Technical Advisory Body;\newline
Two committers elected from CNCF projects &
1 Associate member representative;\newline
1 representative from the Technical Advisory Council \\
\bottomrule
\end{tabular}
\end{table*}

\begin{table*}[t]
\centering
\caption{Technical Committee (or equivalent) representation per membership tier in the
AAIF, CNCF, and PyTorch Foundation. Sources: AAIF TC Charter, CNCF Charter, PyTorch Charter.}
\label{tab:tc-representation}
\begin{threeparttable}
\begin{tabular}{L{1.8cm} L{2.5cm} L{2.5cm} L{3cm} L{2.7cm}}
\toprule
\multicolumn{1}{C{1.8cm}}{\textbf{Membership Tier}} &
\multicolumn{1}{C{2.5cm}}{\textbf{AAIF: Technical Committee (TC)}} &
\multicolumn{1}{C{2.5cm}}{\textbf{CNCF: Technical Oversight Committee (TOC)}} &
\multicolumn{1}{C{3cm}}{\textbf{CNCF: End User Technical Advisory Body (End User TAB)}} &
\multicolumn{1}{C{2.7cm}}{\textbf{PyTorch: Technical Advisory Council (TAC)}} \\
\midrule
\textbf{Platinum} &
1 representative from each member &
N/A &
1 representative from each platinum \textit{end user} member\tnote{a} &
1 representative from each member \\
\addlinespace
\textbf{Gold} &
None &
N/A &
1 representative per 5 gold \textit{end user} members, up to a maximum of 3 &
1 representative per 3 members \\
\addlinespace
\textbf{Silver} &
None &
N/A &
1 representative per 10 silver \textit{end user} members, up to a maximum of 3 &
None \\
\addlinespace
\textbf{Other} &
None &
Governing Board elects 6;\newline
End user Members elect 2;\newline
Non-sandbox project maintainers elect 1;\newline
Existing TOC Members elect a further 2 &
At least 3 by end users at large;\newline
1 or 2 TOC members (each of whom must be employed by an end user member);\newline
1 or more additional representatives from end user members &
One representative from each `TAC project' \\
\bottomrule
\end{tabular}
\begin{tablenotes}
\item[a] An `end user' member of the CNCF is an organisation determined by the Linux Foundation that \textit{``utilizes cloud native technologies internally to build, deploy, and operate software applications, systems, or services''}, explicitly excluding companies that \textit{``engage in the direct commercial distribution, support, or sale of software generated by CNCF projects''}. Source: \url{https://www.cncf.io/enduser/}. Accessed 6/10/2026.
\end{tablenotes}
\end{threeparttable}
\end{table*}

\subsection{Comparison and Discussion}
\label{sec:GovernanceDiscussion}
Given this overview of both IETF's and AAIF/MCP's governance structures, the clearest difference is between the ways in which senior leadership positions are filled. Whereas all but one member of AAIF leadership are representatives of a permanent set of eight private companies and MCP leadership is by invitation only (and currently consists wholly of employees from those same eight companies), a large number of IETF leadership are elected by a random draw of engaged, rank-and-file contributors. Furthermore, while AAIF and MCP contain no pathways through which unpopular leadership can be removed by the community, IETF leadership largely serves two-year fixed terms, and can be removed by a similarly random selection of contributors. Overall, while employees of private companies are represented in both IETF and AAIF/MCP leadership, the differences in governance structures mean that those engaged in the former are able to be held accountable to their actions by the broader community in ways which aren't seen in AAIF/MCP.

It may be argued that the comparison is invalid: the IETF and AAIF are products of very different circumstances, working in different contexts. One could also argue that the IETF is a unique and distinctive organisation that is not without its flaws, and therefore, aiming to replicate it is misguided. However, we do not argue that governance structures employed by the IETF should be reproduced exactly in the case of the AAIF or MCP. Rather, it provides an informative demonstration of potential frameworks for promoting openness and participation in protocol development. Furthermore, when comparing the AAIF to other LF directed funds, such as CNCF and the PyTorch Foundation, a number of noteworthy differences are apparent -- including the differences in GB and TC representation referenced above. First, the `platinum cap' observed in the AAIF above is not found in other directed funds. Second, the AAIF charter specifies a `startup period' lasting three years within which amendments to the charter, the adoption of new rules or policies, an increase to the `platinum cap', and the approval of additional projects can be vetoed by platinum members that have contributed projects to the AAIF.\footnote{\url{https://github.com/aaif/foundation/blob/main/foundation-charter.pdf}, 5(e). Accessed 6/10/2026.} This `startup period' also appears to be unique to the AAIF. Finally, it is worth pointing out a number of features of the CNCF that are absent from the AAIF. The CNCF charter includes a `values' section, that explicitly lists `neutrality' as a core value, stating that the CNCF `avoid[s] influence, biased behavior, ``pay-to-play'' decision-making, and other activities that favor one project, individual, or organization over another' (CNCF Charter, §3(c)). The CNCF also allows for members to be designated as `end user' organisations, which explicitly excludes companies developing or distributing CNCF technologies, with a separate board (the `End User Technical Advisory Body') reserved solely for such members (CNCF charter, §7-8).

When it comes to MCP itself, stewardship explicitly takes the form of a `Benevolent Dictator For Life' (BDFL) model with the two lead maintainers holding ultimate decision-making authority. Since MCP's donation to the AAIF, its governance structure has become increasingly formalised, including the provision that only lead maintainers are able to propose amendments to the governance structure \citep{soria_parra_governance_2025}. At time of writing, there is also a notable amount of overlap between the MCP core maintainers, the AAIF GB, and the AAIF TC. In particular, an MCP lead maintainer is the chair of the TC, the TC co-chair is an MCP core maintainer, and a third individual is simultaneously an MCP core maintainer and a representative on both the TC and GB.\footnote{ \url{https://modelcontextprotocol.io/community/governance}; \url{https://aaif.io/board/}; \url{https://aaif.io/tc/}. All accessed 6/10/2026.} In contrast, Kubernetes, a project under the CNCF, has a seven-member steering committee elected to two-year terms by active contributors to the project.\footnote{See \url{https://github.com/kubernetes/steering/blob/main/charter.md}; \url{https://github.com/kubernetes/steering/blob/main/elections.md}; \url{https://github.com/kubernetes/community/tree/main/elections/steering/2026}. All accessed 6/10/2026.} A closer comparison is PyTorch, which, while also adopting a BDFL structure, includes a provision such that a super-majority of core maintainers can choose to remove the lead maintainer.\footnote{\url{https://docs.pytorch.org/docs/stable/community/governance.html}. Accessed 6/10/2026.}

The picture that emerges is one in which numerous design choices, each of which might seem inconsequential in isolation, collectively centralise decision-making authority in a small number of tech companies and their employees, with limited routes to community accountability or representation. This pattern is not unprecedented in the AI industry: \citet{widder_open_2023} observe that
\begin{quote}
    some companies have moved to embrace `open' AI as a mechanism to entrench dominance, using the rhetoric of `open' AI to expand market power while investing in `open' AI efforts in ways that allow them to set standards of development while benefitting from the free labor of open source contributors.
\end{quote}
In this context, the donation of MCP to AAIF warrants scrutiny. 
The donation does not affect MCP's BDFL model and lack of mechanism for removing lead maintainers, the platinum cap restricts engagement with the AAIF at the highest levels, and the three-year startup period in the AAIF charter means that structural amendments to the foundation can be vetoed by any of Anthropic, Block, or OpenAI.
This is noteworthy given the absence of any acknowledgement or policy regarding potential conflicts of interest or a `values' section of the charter -- as seen for CNCF.

While we do not make any claim that this authority is being abused, we note that it does create the possibility of unaccountable use of this power in ways that may not be in the broader community's interest. Concretely, lead maintainers could steer the protocol in a direction that favours their own models or infrastructure, or block functionality that benefits competitors. Less dramatically, leadership could refuse to implement important functionality if it is not aligned with the interest of their employer(s), including some security measures.

Potential responses to these concerns include that a concentrated power structure is necessary to ensure the stability and coherence of MCP governance and strategy while the project is rapidly scaling, and that this will be relaxed once the rate of development slows. However, other than the aforementioned `startup period' in the AAIF charter, there is no mention of any governance structures being time-limited. Furthermore, actions taken by leadership, such as restricting the right to propose governance amendments to lead maintainers, mean that any decision to increase accountability and representation can only be made by two people \citep{soria_parra_governance_2025}. Alternatively, one might argue that contributors and community members that are unhappy with the governance structures are free to work on an alternative project, or fork the repository to proceed in a different direction. However, while such defection is theoretically possible, given MCP's current scale and extent of corporate adoption, direct competition is likely infeasible in practice, and is further hindered by the fact that many potential corporate backers of an alternative protocol already support MCP through the AAIF. Additionally, this argument assumes that only contributors have a valid interest in how MCP is governed. However, given that MCP is often referred to by its own stewards as becoming `critical infrastructure for AI' \citep{nisha_how_2025}, it is not unreasonable for individuals who are not involved in its development to be concerned with how it is governed.

Ultimately, there is a question of legitimacy when it comes to who gets to control `critical infrastructure', whether relating to the internet, AI, or otherwise. As highlighted by \citet{denardis_protocol_2009}, `if technical standards make public interest decisions, the questions of who sets technical standards and how they set them are highly relevant.' While the IETF -- a technocratic, small (on the order of thousands), and unrepresentative subset of the population that's responsible for infrastructure used by billions -- may not be a paragon of democratic legitimacy, its adoption of unique governance mechanisms aimed at increasing participation and accountability can provide inspiration for individuals and organisations leading the development of AI protocols. Should MCP achieve its stated aim of becoming `critical infrastructure for AI' with the current governance structures in place, this would represent a significant concentration of authority over a technology with far-reaching implications, grounded primarily in the speed of its adoption.

\section{Discussion}
\label{sec:DiscussionConclusion}
Overall, the preceding two sections combine to paint a cautionary picture. If steps are not taken early to implement security, the ecosystem may burden itself with a persistent security deficit with ever-increasing costs of retrofitting. This security deficit may emerge alongside entrenched governance structures that preserve corporations' primacy with regards to long-term decision-making, while public statements emphasise community input and participation.

Notably, these conclusions may be in tension: it is plausible that the more concentrated power structure of MCP/AAIF is beneficial or even necessary for ensuring the security of MCP, since key design features that enhance security could hypothetically be prioritised from the top down. The concepts of input and output legitimacy, introduced by \citet{scharpf_governing_1999} in the context of the European Union, and \citet{schmidt_democracy_2013}'s extension to `throughput' legitimacy may be a relevant framing here, having been applied to internet governance in \citep{palladino_legitimacy_2021}. Input legitimacy refers to the extent to which governance is \textit{`by the people'} -- i.e., that subjects' views, preferences, and beliefs are taken as inputs to the governance processes. In contrast, output legitimacy refers to the extent to which governance is \textit{`for the people'} -- i.e., that the actions or decisions taken as a result of a governance process serve subjects' interests. Throughput legitimacy adds an intermediate layer, concerning the processes by which governance decisions are made, for example, whether they are transparent and accountable. In this paper, while we make no assessment of MCP or AAIF's output legitimacy, instead observing multiple features of their input and throughput legitimacy. Whilst MCP's openness to contributions from the broader community may in some cases lead to increased input legitimacy, AAIF's cap on the number of platinum members and restricted representation on its governing board may decrease input legitimacy. Regarding throughput legitimacy, MCP's transparency on account of it being developed open source is counteracted by limited mechanisms for holding leadership accountable for design decisions. \citet{schmidt_democracy_2013} argues further that a lack of throughput legitimacy can serve to `delegitimate both input and output'. In the case of MCP, this would imply that output legitimacy, perhaps in the form of increased security, could be undermined by an illegitimate development process.

In this paper we have largely focused on MCP as an archetypal case study for AI agent protocols due to its status as the most widely adopted AI agent protocol. However, the broad effects observed here concerning the practical and economic challenges of retrofitting security, and the nature of governance and representation, are generalisable regardless of the specific project in question. Should MCP become defunct or outcompeted by an alternative, we believe that whatever project takes its place will have to contend with similar security considerations. While the discussion of governance structures in Sections \ref{sec:AAIF} and \ref{sec:GovernanceDiscussion} is specific to MCP and the AAIF, we focus on these structures because of MCP's rapid uptake among users and business -- making the question of how it is governed particularly consequential.

Our conclusions carry several practical implications. First, policymakers should accept that de facto standardisation can come from sources outside of traditional SDOs and policymaking institutions. NIST's recent AI Agent Standards Initiative, which actively engages with industry-led protocols development reflects a growing acknowledgement of this reality.\footnote{\url{https://www.nist.gov/news-events/news/2026/02/announcing-ai-agent-standards-initiative-interoperable-and-secure}. Accessed 6/10/2026.} Ignoring such developments while following traditional formal standards setting processes risks being outcompeted by more nimble, industry-led efforts. Second, claims of projects being `open, neutral and community-driven' should be approached with a critical eye, and judgements should be based on how such projects operate in practice, rather than public statements. Third, the observed difficulty with which security was retrofitted to internet protocols should be taken as a lower bound for that which could be faced by the ecosystem of AI agent protocols. Finally, there is an important unanswered question of designing protocol governance structures that are simultaneously nimble and maintain sufficient representation and accountability for community interests, including security, to be taken seriously. We encourage protocol developers, policymakers, and affected communities to proactively engage in debate and experimentation on this issue.

\begin{acks}
The authors would like to thank Alan Chan, Maike Osborne, Amin Oueslati, Patricia Paskov, Huw Roberts, Chris Schmitz, Merlin Stein, and Marta Ziosi for comments and feedback.
\end{acks}



\bibliographystyle{ACM-Reference-Format}
\bibliography{protocols}

@incollection{russell_internet_2014,
	edition = {1},
	title = {The {Internet} and the {Advantages} of {Autocratic} {Design}, 1970s-1990s},
	copyright = {https://www.cambridge.org/core/terms},
	isbn = {978-1-107-03919-3 978-1-107-61204-4 978-1-139-85655-3},
	url = {https://www.cambridge.org/core/product/identifier/9781139856553/type/book},
	doi = {10.1017/CBO9781139856553},
	language = {en},
	urldate = {2026-03-09},
	booktitle = {Open {Standards} and the {Digital} {Age}: {History}, {Ideology}, and {Networks}},
	publisher = {Cambridge University Press},
	author = {Russell, Andrew L.},
	month = apr,
	year = {2014},
}

@misc{scott_policy_2015,
	title = {Policy to {Require} {Secure} {Connections} across {Federal} {Websites} and {Web} {Services}},
	url = {https://obamawhitehouse.archives.gov/sites/default/files/omb/memoranda/2015/m-15-13.pdf},
	language = {en},
	author = {Scott, Tony},
	month = jun,
	year = {2015},
	note = {https://obamawhitehouse.archives.gov/sites/default/files/omb/memoranda/2015/m-15-13.pdf},
}

@misc{crocker_origins_1987,
	type = {Request for {Comments}},
	title = {The {Origins} of {RFCs}},
	url = {https://datatracker.ietf.org/doc/rfc1000},
	doi = {10.17487/RFC1000},
	urldate = {2026-03-09},
	publisher = {Internet Engineering Task Force},
	author = {Crocker, Stephen D.},
	month = aug,
	year = {1987},
	note = {RFC 1000, https://datatracker.ietf.org/doc/rfc1000},
}

@misc{widder_open_2023,
	address = {Rochester, NY},
	type = {{SSRN} {Scholarly} {Paper}},
	title = {Open ({For} {Business}): {Big} {Tech}, {Concentrated} {Power}, and the {Political} {Economy} of {Open} {AI}},
	shorttitle = {Open ({For} {Business})},
	url = {https://papers.ssrn.com/abstract=4543807},
	doi = {10.2139/ssrn.4543807},
	language = {en},
	urldate = {2026-03-20},
	publisher = {Social Science Research Network},
	author = {Widder, David Gray and West, Sarah and Whittaker, Meredith},
	month = aug,
	year = {2023},
}

@book{denardis_protocol_2009,
	address = {Cambridge, UNITED STATES},
	title = {Protocol {Politics}: {The} {Globalization} of {Internet} {Governance}},
	isbn = {978-0-262-25873-9},
	shorttitle = {Protocol {Politics}},
	url = {http://ebookcentral.proquest.com/lib/oxford/detail.action?docID=3339043},
	urldate = {2026-03-24},
	publisher = {MIT Press},
	author = {DeNardis, Laura},
	year = {2009},
}

@article{bellovin_security_1989,
	title = {Security problems in the {TCP}/{IP} protocol suite},
	volume = {19},
	issn = {0146-4833},
	url = {https://dl.acm.org/doi/10.1145/378444.378449},
	doi = {10.1145/378444.378449},
	number = {2},
	urldate = {2026-03-25},
	journal = {SIGCOMM Comput. Commun. Rev.},
	author = {Bellovin, S. M.},
	month = apr,
	year = {1989},
	pages = {32--48},
}

@misc{errico_securing_2025,
	title = {Securing the {Model} {Context} {Protocol} ({MCP}): {Risks}, {Controls}, and {Governance}},
	shorttitle = {Securing the {Model} {Context} {Protocol} ({MCP})},
	url = {http://arxiv.org/abs/2511.20920},
	doi = {10.48550/arXiv.2511.20920},
	urldate = {2026-03-25},
	publisher = {arXiv},
	author = {Errico, Herman and Ngiam, Jiquan and Sojan, Shanita},
	month = nov,
	year = {2025},
	note = {arXiv:2511.20920 [cs]},
}

@misc{zhao_when_2025,
	title = {When {MCP} {Servers} {Attack}: {Taxonomy}, {Feasibility}, and {Mitigation}},
	shorttitle = {When {MCP} {Servers} {Attack}},
	url = {http://arxiv.org/abs/2509.24272},
	doi = {10.48550/arXiv.2509.24272},
	urldate = {2026-03-25},
	publisher = {arXiv},
	author = {Zhao, Weibo and Liu, Jiahao and Ruan, Bonan and Li, Shaofei and Liang, Zhenkai},
	month = sep,
	year = {2025},
	note = {arXiv:2509.24272 [cs]},
}

@misc{anbiaee_security_2026,
	title = {Security {Threat} {Modeling} for {Emerging} {AI}-{Agent} {Protocols}: {A} {Comparative} {Analysis} of {MCP}, {A2A}, {Agora}, and {ANP}},
	shorttitle = {Security {Threat} {Modeling} for {Emerging} {AI}-{Agent} {Protocols}},
	url = {http://arxiv.org/abs/2602.11327},
	doi = {10.48550/arXiv.2602.11327},
	urldate = {2026-03-25},
	publisher = {arXiv},
	author = {Anbiaee, Zeynab and Rabbani, Mahdi and Mirani, Mansur and Piya, Gunjan and Opushnyev, Igor and Ghorbani, Ali and Dadkhah, Sajjad},
	month = feb,
	year = {2026},
	note = {arXiv:2602.11327 [cs]},
}

@misc{radosevich_mcp_2025,
	title = {{MCP} {Safety} {Audit}: {LLMs} with the {Model} {Context} {Protocol} {Allow} {Major} {Security} {Exploits}},
	shorttitle = {{MCP} {Safety} {Audit}},
	url = {http://arxiv.org/abs/2504.03767},
	doi = {10.48550/arXiv.2504.03767},
	urldate = {2026-03-25},
	publisher = {arXiv},
	author = {Radosevich, Brandon and Halloran, John},
	month = apr,
	year = {2025},
	note = {arXiv:2504.03767 [cs]},
}

@article{hou_model_2026,
	title = {Model {Context} {Protocol} ({MCP}): {Landscape}, {Security} {Threats}, and {Future} {Research} {Directions}},
	issn = {1049-331X},
	shorttitle = {Model {Context} {Protocol} ({MCP})},
	url = {https://dl.acm.org/doi/10.1145/3796519},
	doi = {10.1145/3796519},
	urldate = {2026-03-25},
	journal = {ACM Trans. Softw. Eng. Methodol.},
	author = {Hou, Xinyi and Zhao, Yanjie and Wang, Shenao and Wang, Haoyu},
	month = feb,
	year = {2026},
}

@misc{guo_systematic_2025,
	title = {Systematic {Analysis} of {MCP} {Security}},
	url = {http://arxiv.org/abs/2508.12538},
	doi = {10.48550/arXiv.2508.12538},
	urldate = {2026-03-25},
	publisher = {arXiv},
	author = {Guo, Yongjian and Liu, Puzhuo and Ma, Wanlun and Deng, Zehang and Zhu, Xiaogang and Di, Peng and Xiao, Xi and Wen, Sheng},
	month = aug,
	year = {2025},
	note = {arXiv:2508.12538 [cs]},
}

@incollection{harcourt_internal_2020,
	title = {Internal {Governance} of the {IETF}, {W3C}, {OASIS}, and {IEEE}: {Structure}, {Decision}-making, and {Internationalization}},
	isbn = {978-0-19-884152-4},
	shorttitle = {Internal {Governance} of the {IETF}, {W3C}, {OASIS}, and {IEEE}},
	url = {https://doi.org/10.1093/oso/9780198841524.003.0003},
	doi = {10.1093/oso/9780198841524.003.0003},
	urldate = {2026-03-26},
	booktitle = {Global {Standard} {Setting} in {Internet} {Governance}},
	publisher = {Oxford University Press},
	author = {Harcourt, Alison and Christou, George and Simpson, Seamus},
	editor = {Harcourt, Alison and Christou, George and Simpson, Seamus},
	month = jan,
	year = {2020},
	pages = {0},
}

@misc{stein_how_2026,
	title = {How are {AI} agents used? {Evidence} from 177,000 {MCP} tools},
	shorttitle = {How are {AI} agents used?},
	url = {http://arxiv.org/abs/2603.23802},
	doi = {10.48550/arXiv.2603.23802},
	urldate = {2026-03-26},
	publisher = {arXiv},
	author = {Stein, Merlin},
	month = mar,
	year = {2026},
	note = {arXiv:2603.23802 [cs]},
}

@incollection{denardis_history_2007,
	title = {A history of internet security},
	isbn = {978-0-444-51608-4},
	url = {https://www-sciencedirect-com.ezproxy-prd.bodleian.ox.ac.uk/science/chapter/edited-volume/pii/B9780444516084500250},
	doi = {10.1016/B978-044451608-4/50025-0},
	language = {en-US},
	urldate = {2026-04-07},
	booktitle = {The {History} of {Information} {Security}: {A} {Comprehensive} {Handbook}},
	publisher = {Elsevier Science B.V.},
	author = {DeNardis, Laura},
	editor = {De Leeuw, Karl and Bergstra, Jan},
	month = jan,
	year = {2007},
	doi = {10.1016/B978-044451608-4/50025-0},
	pages = {681--704},
}

@misc{bradner_internet_1996,
	type = {Request for {Comments}},
	title = {The {Internet} {Standards} {Process} – {Revision} 3},
	url = {https://datatracker.ietf.org/doc/rfc2026},
	doi = {10.17487/RFC2026},
	urldate = {2026-04-09},
	publisher = {Internet Engineering Task Force},
	author = {Bradner, Scott O.},
	month = oct,
	year = {1996},
	note = {RFC 2026, https://datatracker.ietf.org/doc/rfc2026},
}

@misc{kucherawy_iab_2020,
	type = {Request for {Comments}},
	title = {{IAB}, {IESG}, {IETF} {Trust}, and {IETF} {LLC} {Selection}, {Confirmation}, and {Recall} {Process}: {Operation} of the {IETF} {Nominating} and {Recall} {Committees}},
	shorttitle = {{IAB}, {IESG}, {IETF} {Trust}, and {IETF} {LLC} {Selection}, {Confirmation}, and {Recall} {Process}},
	url = {https://datatracker.ietf.org/doc/rfc8713},
	doi = {10.17487/RFC8713},
	urldate = {2026-04-09},
	publisher = {Internet Engineering Task Force},
	author = {Kucherawy, Murray and Hinden, Bob and Livingood, Jason},
	month = feb,
	year = {2020},
	note = {RFC 8713, https://datatracker.ietf.org/doc/rfc8713},
}

@misc{salz_entities_2022,
	type = {Request for {Comments}},
	title = {Entities {Involved} in the {IETF} {Standards} {Process}},
	url = {https://datatracker.ietf.org/doc/rfc9281},
	doi = {10.17487/RFC9281},
	urldate = {2026-04-09},
	publisher = {Internet Engineering Task Force},
	author = {Salz, Rich},
	month = jun,
	year = {2022},
	note = {RFC 9281, https://datatracker.ietf.org/doc/rfc9281},
}

@misc{bradner_ietf_1998,
	type = {Request for {Comments}},
	title = {{IETF} {Working} {Group} {Guidelines} and {Procedures}},
	url = {https://datatracker.ietf.org/doc/rfc2418},
	doi = {10.17487/RFC2418},
	urldate = {2026-04-09},
	publisher = {Internet Engineering Task Force},
	author = {Bradner, Scott O.},
	month = sep,
	year = {1998},
	note = {RFC 2418, https://datatracker.ietf.org/doc/rfc2418},
}

@misc{alvestrand_mission_2004,
	type = {Request for {Comments}},
	title = {A {Mission} {Statement} for the {IETF}},
	url = {https://datatracker.ietf.org/doc/rfc3935},
	doi = {10.17487/RFC3935},
	urldate = {2026-04-09},
	publisher = {Internet Engineering Task Force},
	author = {Alvestrand, Harald T.},
	month = oct,
	year = {2004},
	note = {RFC 3935, https://datatracker.ietf.org/doc/rfc3935},
}

@misc{bradner_intellectual_2017,
	type = {Request for {Comments}},
	title = {Intellectual {Property} {Rights} in {IETF} {Technology}},
	url = {https://datatracker.ietf.org/doc/rfc8179},
	doi = {10.17487/RFC8179},
	urldate = {2026-04-09},
	publisher = {Internet Engineering Task Force},
	author = {Bradner, Scott O. and Contreras, Jorge},
	month = may,
	year = {2017},
	note = {RFC 8179, https://datatracker.ietf.org/doc/rfc8179},
}

@misc{resnick_consensus_2014,
	type = {Request for {Comments}},
	title = {On {Consensus} and {Humming} in the {IETF}},
	url = {https://datatracker.ietf.org/doc/rfc7282},
	doi = {10.17487/RFC7282},
	urldate = {2026-04-09},
	publisher = {Internet Engineering Task Force},
	author = {Resnick, Pete},
	month = jun,
	year = {2014},
	note = {RFC 7282, https://datatracker.ietf.org/doc/rfc7282},
}

@misc{eastlake_3rd_publicly_2004,
	type = {Request for {Comments}},
	title = {Publicly {Verifiable} {Nominations} {Committee} ({NomCom}) {Random} {Selection}},
	url = {https://datatracker.ietf.org/doc/rfc3797},
	doi = {10.17487/RFC3797},
	urldate = {2026-04-09},
	publisher = {Internet Engineering Task Force},
	author = {Eastlake 3rd, Donald E.},
	month = jun,
	year = {2004},
	note = {RFC 3797, https://datatracker.ietf.org/doc/rfc3797},
}

@misc{duke_nominating_2023,
	type = {Request for {Comments}},
	title = {Nominating {Committee} {Eligibility}},
	url = {https://datatracker.ietf.org/doc/rfc9389},
	doi = {10.17487/RFC9389},
	urldate = {2026-04-09},
	publisher = {Internet Engineering Task Force},
	author = {Duke, Martin},
	month = apr,
	year = {2023},
	note = {RFC 9389, https://datatracker.ietf.org/doc/rfc9389},
}

@misc{franklin_ai_2026,
	address = {Rochester, NY},
	type = {{SSRN} {Scholarly} {Paper}},
	title = {{AI} {Agent} {Traps}},
	url = {https://papers.ssrn.com/abstract=6372438},
	doi = {10.2139/ssrn.6372438},
	language = {en},
	urldate = {2026-04-21},
	publisher = {Social Science Research Network},
	author = {Franklin, Matija and Tomašev, Nenad and Jacobs, Julian and Leibo, Joel Z. and Osindero, Simon},
	month = mar,
	year = {2026},
}

@misc{croce_trivial_2025,
	title = {Trivial {Trojans}: {How} {Minimal} {MCP} {Servers} {Enable} {Cross}-{Tool} {Exfiltration} of {Sensitive} {Data}},
	shorttitle = {Trivial {Trojans}},
	url = {http://arxiv.org/abs/2507.19880},
	doi = {10.48550/arXiv.2507.19880},
	urldate = {2026-04-21},
	publisher = {arXiv},
	author = {Croce, Nicola and South, Tobin},
	month = jul,
	year = {2025},
	note = {arXiv:2507.19880 [cs]},
}

@article{anderson_economics_2006,
	title = {The {Economics} of {Information} {Security}},
	volume = {314},
	issn = {0036-8075},
	url = {https://www.jstor.org/stable/20031627},
	number = {5799},
	urldate = {2026-04-21},
	journal = {Science},
	publisher = {American Association for the Advancement of Science},
	author = {Anderson, Ross and Moore, Tyler},
	year = {2006},
	pages = {610--613},
}

@inproceedings{ozment_bootstrapping_2006,
	title = {Bootstrapping the {Adoption} of {Internet} {Security} {Protocols}},
	url = {https://www.econinfosec.org/archive/weis2006/docs/46.pdf},
	language = {en},
	booktitle = {Proceedings of the {Fifth} {Workshop} on the {Economics} of {Information} {Security}},
	author = {Ozment, Andy and Schechter, Stuart E},
	month = jun,
	year = {2006},
}

@article{bauer_cybersecurity_2009,
	title = {Cybersecurity: {Stakeholder} incentives, externalities, and policy options},
	volume = {33},
	issn = {0308-5961},
	shorttitle = {Cybersecurity},
	url = {https://www.sciencedirect.com/science/article/pii/S0308596109000986},
	doi = {10.1016/j.telpol.2009.09.001},
	number = {10},
	urldate = {2026-04-21},
	journal = {Telecommunications Policy},
	author = {Bauer, Johannes M. and van Eeten, Michel J. G.},
	month = nov,
	year = {2009},
	pages = {706--719},
}

@article{gordon_externalities_2015,
	title = {Externalities and the {Magnitude} of {Cyber} {Security} {Underinvestment} by {Private} {Sector} {Firms}: {A} {Modification} of the {Gordon}-{Loeb} {Model}},
	volume = {6},
	copyright = {http://creativecommons.org/licenses/by/4.0/},
	shorttitle = {Externalities and the {Magnitude} of {Cyber} {Security} {Underinvestment} by {Private} {Sector} {Firms}},
	url = {https://www.scirp.org/journal/paperinformation?paperid=52952},
	doi = {10.4236/jis.2015.61003},
	language = {en},
	number = {1},
	urldate = {2026-04-21},
	journal = {Journal of Information Security},
	publisher = {Scientific Research Publishing},
	author = {Gordon, Lawrence A. and Loeb, Martin P. and Lucyshyn, William and Zhou, Lei},
	month = jan,
	year = {2015},
	pages = {24--30},
}

@article{kunreuther_interdependent_2003,
	title = {Interdependent {Security}},
	volume = {26},
	issn = {0895-5646},
	url = {https://www.jstor.org/stable/41755017},
	number = {2/3},
	urldate = {2026-04-21},
	journal = {Journal of Risk and Uncertainty},
	publisher = {Springer},
	author = {Kunreuther, Howard and Heal, Geoffrey},
	year = {2003},
	pages = {231--249},
}

@incollection{camp_system_2004,
	address = {Boston},
	title = {System {Reliability} and {Free} {Riding}},
	volume = {12},
	isbn = {978-1-4020-8089-0},
	url = {http://link.springer.com/10.1007/1-4020-8090-5_1},
	doi = {10.1007/1-4020-8090-5_1},
	language = {en},
	urldate = {2026-04-21},
	booktitle = {Economics of {Information} {Security}},
	publisher = {Kluwer Academic Publishers},
	author = {Varian, Hal},
	editor = {Camp, L. Jean and Lewis, Stephen},
	year = {2004},
	note = {Series Title: Advances in Information Security},
	pages = {1--15},
}

@article{chan_infrastructure_2025,
	title = {Infrastructure for {AI} {Agents}},
	issn = {2835-8856},
	url = {https://openreview.net/forum?id=Ckh17xN2R2},
	language = {en},
	urldate = {2026-04-22},
	journal = {Transactions on Machine Learning Research},
	author = {Chan, Alan and Wei, Kevin and Huang, Sihao and Rajkumar, Nitarshan and Perrier, Elija and Lazar, Seth and Hadfield, Gillian K. and Anderljung, Markus},
	month = feb,
	year = {2025},
}

@misc{yang_survey_2025,
	title = {A {Survey} of {AI} {Agent} {Protocols}},
	url = {https://arxiv.org/abs/2504.16736v3},
	language = {en},
	urldate = {2026-04-22},
	publisher = {arXiv},
	author = {Yang, Yingxuan and Chai, Huacan and Song, Yuanyi and Qi, Siyuan and Wen, Muning and Li, Ning and Liao, Junwei and Hu, Haoyi and Lin, Jianghao and Chang, Gaowei and Liu, Weiwen and Wen, Ying and Yu, Yong and Zhang, Weinan},
	month = apr,
	year = {2025},
	note = {arXiv:2504.16716v3 [cs]},
}

@techreport{booth_accelerating_2026,
	title = {Accelerating the {Adoption} of {Software} and {AI} {Agent} {Identity} and {Authorization}},
	language = {en},
	institution = {NIST, NCCoE},
	author = {Booth, Harold and Fisher, Bill and Galluzzo, Ryan and Roberts, Joshua},
	month = feb,
	year = {2026},
}

@misc{surapaneni_announcing_2025,
	title = {Announcing the {Agent2Agent} {Protocol} ({A2A})},
	url = {https://developers.googleblog.com/en/a2a-a-new-era-of-agent-interoperability/},
	language = {en},
	urldate = {2026-04-23},
	journal = {Google Developers Blog},
	author = {Surapaneni, Rao and Jha, Miku and Vakoc, Michael and Segal, Todd},
	month = apr,
	year = {2025},
	note = {https://developers.googleblog.com/en/a2a-a-new-era-of-agent-interoperability/},
}

@misc{martineau_open-source_2025,
	title = {An open-source protocol for {AI} agents to interact},
	copyright = {© Copyright IBM Corp. 2021},
	url = {https://research.ibm.com/blog/agent-communication-protocol-ai},
	language = {en-US},
	urldate = {2026-04-23},
	journal = {IBM Research},
	author = {Martineau, Kim},
	month = may,
	year = {2025},
	note = {https://research.ibm.com/blog/agent-communication-protocol-ai},
}

@misc{ehtesham_survey_2025,
	title = {A survey of agent interoperability protocols: {Model} {Context} {Protocol} ({MCP}), {Agent} {Communication} {Protocol} ({ACP}), {Agent}-to-{Agent} {Protocol} ({A2A}), and {Agent} {Network} {Protocol} ({ANP})},
	shorttitle = {A survey of agent interoperability protocols},
	url = {http://arxiv.org/abs/2505.02279},
	doi = {10.48550/arXiv.2505.02279},
	urldate = {2026-04-23},
	publisher = {arXiv},
	author = {Ehtesham, Abul and Singh, Aditi and Gupta, Gaurav Kumar and Kumar, Saket},
	month = may,
	year = {2025},
	note = {arXiv:2505.02279 [cs]},
}

@article{eisenberg_cornell_1989,
	title = {The {Cornell} commission: on {Morris} and the worm},
	volume = {32},
	issn = {0001-0782},
	shorttitle = {The {Cornell} commission},
	url = {https://dl.acm.org/doi/10.1145/63526.63530},
	doi = {10.1145/63526.63530},
	number = {6},
	urldate = {2026-04-23},
	journal = {Commun. ACM},
	author = {Eisenberg, T. and Gries, D. and Hartmanis, J. and Holcomb, D. and Lynn, M. S. and Santoro, T.},
	month = jun,
	year = {1989},
	pages = {706--709},
}

@inproceedings{aas_lets_2019,
	address = {New York, NY, USA},
	series = {{CCS} '19},
	title = {Let's {Encrypt}: {An} {Automated} {Certificate} {Authority} to {Encrypt} the {Entire} {Web}},
	isbn = {978-1-4503-6747-9},
	shorttitle = {Let's {Encrypt}},
	url = {https://dl.acm.org/doi/10.1145/3319535.3363192},
	doi = {10.1145/3319535.3363192},
	urldate = {2026-04-23},
	booktitle = {Proceedings of the 2019 {ACM} {SIGSAC} {Conference} on {Computer} and {Communications} {Security}},
	publisher = {Association for Computing Machinery},
	author = {Aas, Josh and Barnes, Richard and Case, Benton and Durumeric, Zakir and Eckersley, Peter and Flores-López, Alan and Halderman, J. Alex and Hoffman-Andrews, Jacob and Kasten, James and Rescorla, Eric and Schoen, Seth and Warren, Brad},
	month = nov,
	year = {2019},
	pages = {2473--2487},
}

@inproceedings{felt_measuring_2017,
	title = {Measuring {HTTPS} {Adoption} on the {Web}},
	isbn = {978-1-931971-40-9},
	url = {https://www.usenix.org/conference/usenixsecurity17/technical-sessions/presentation/felt},
	language = {en},
	urldate = {2026-04-23},
	author = {Felt, Adrienne Porter and Barnes, Richard and King, April and Palmer, Chris and Bentzel, Chris and Tabriz, Parisa},
	year = {2017},
	pages = {1323--1338},
}

@inproceedings{naylor_cost_2014,
	address = {New York, NY, USA},
	series = {{CoNEXT} '14},
	title = {The {Cost} of the "{S}" in {HTTPS}},
	isbn = {978-1-4503-3279-8},
	url = {https://dl.acm.org/doi/10.1145/2674005.2674991},
	doi = {10.1145/2674005.2674991},
	urldate = {2026-04-23},
	booktitle = {Proceedings of the 10th {ACM} {International} on {Conference} on emerging {Networking} {Experiments} and {Technologies}},
	publisher = {Association for Computing Machinery},
	author = {Naylor, David and Finamore, Alessandro and Leontiadis, Ilias and Grunenberger, Yan and Mellia, Marco and Munafò, Maurizio and Papagiannaki, Konstantina and Steenkiste, Peter},
	month = dec,
	year = {2014},
	pages = {133--140},
}

@inproceedings{felt_rethinking_2016,
	title = {Rethinking {Connection} {Security} {Indicators}},
	isbn = {978-1-931971-31-7},
	url = {https://www.usenix.org/conference/soups2016/technical-sessions/presentation/porter-felt},
	language = {en},
	urldate = {2026-04-23},
	author = {Felt, Adrienne Porter and Reeder, Robert W. and Ainslie, Alex and Harris, Helen and Walker, Max and Thompson, Christopher and Acer, Mustafa Embre and Morant, Elisabeth and Consolvo, Sunny},
	year = {2016},
	pages = {1--14},
}

@inproceedings{krombholz_i_2017,
	title = {"{I} {Have} {No} {Idea} {What} {I}'m {Doing}" - {On} the {Usability} of {Deploying} {HTTPS}},
	isbn = {978-1-931971-40-9},
	url = {https://www.usenix.org/conference/usenixsecurity17/technical-sessions/presentation/krombholz},
	language = {en},
	urldate = {2026-04-23},
	author = {Krombholz, Katharina and Mayer, Wilfried and Schmiedecker, Martin and Weippl, Edgar},
	year = {2017},
	pages = {1339--1356},
}

@inproceedings{bernhard_usability_2019,
	address = {New York, NY, USA},
	series = {{CHI} '19},
	title = {On the {Usability} of {HTTPS} {Deployment}},
	isbn = {978-1-4503-5970-2},
	url = {https://dl.acm.org/doi/10.1145/3290605.3300540},
	doi = {10.1145/3290605.3300540},
	urldate = {2026-04-23},
	booktitle = {Proceedings of the 2019 {CHI} {Conference} on {Human} {Factors} in {Computing} {Systems}},
	publisher = {Association for Computing Machinery},
	author = {Bernhard, Matthew and Sharman, Jonathan and Acemyan, Claudia Ziegler and Kortum, Philip and Wallach, Dan S. and Halderman, J. Alex},
	month = may,
	year = {2019},
	pages = {1--10},
}

@article{kuerbis_hidden_2020,
	title = {The hidden standards war: economic factors affecting {IPv6} deployment},
	volume = {22},
	issn = {2398-5038},
	shorttitle = {The hidden standards war},
	url = {https://doi.org/10.1108/DPRG-10-2019-0085},
	doi = {10.1108/DPRG-10-2019-0085},
	number = {4},
	urldate = {2026-04-23},
	journal = {Digital Policy, Regulation and Governance},
	author = {Kuerbis, Brenden and Mueller, Milton},
	month = nov,
	year = {2020},
	pages = {333--361},
}

@article{wu_transition_2013,
	title = {Transition from {IPv4} to {IPv6}: {A} {State}-of-the-{Art} {Survey}},
	volume = {15},
	issn = {1553-877X},
	shorttitle = {Transition from {IPv4} to {IPv6}},
	url = {https://ieeexplore.ieee.org/document/6380492},
	doi = {10.1109/SURV.2012.110112.00200},
	number = {3},
	urldate = {2026-04-23},
	journal = {IEEE Communications Surveys \& Tutorials},
	author = {Wu, Peng and Cui, Yong and Wu, Jianping and Liu, Jiangchuan and Metz, Chris},
	year = {2013},
	pages = {1407--1424},
}

@inproceedings{gouda_ru-rpki-ready_2025,
	address = {New York, NY, USA},
	series = {{IMC} '25},
	title = {ru-{RPKI}-ready: the {Road} {Left} to {Full} {ROA} {Adoption}},
	isbn = {979-8-4007-1860-1},
	shorttitle = {ru-{RPKI}-ready},
	url = {https://dl.acm.org/doi/10.1145/3730567.3764452},
	doi = {10.1145/3730567.3764452},
	urldate = {2026-04-23},
	booktitle = {Proceedings of the 2025 {ACM} {Internet} {Measurement} {Conference}},
	publisher = {Association for Computing Machinery},
	author = {Gouda, Deepak and Fontugne, Romain and Testart, Cecilia},
	month = nov,
	year = {2025},
	pages = {415--429},
}

@article{leiner_brief_2009,
	title = {A brief history of the internet},
	volume = {39},
	issn = {0146-4833},
	url = {https://dl.acm.org/doi/10.1145/1629607.1629613},
	doi = {10.1145/1629607.1629613},
	number = {5},
	urldate = {2026-04-23},
	journal = {SIGCOMM Comput. Commun. Rev.},
	author = {Leiner, Barry M. and Cerf, Vinton G. and Clark, David D. and Kahn, Robert E. and Kleinrock, Leonard and Lynch, Daniel C. and Postel, Jon and Roberts, Larry G. and Wolff, Stephen},
	month = oct,
	year = {2009},
	pages = {22--31},
}

@misc{berners-lee_hypertext_1996,
	type = {Request for {Comments}},
	title = {Hypertext {Transfer} {Protocol} – {HTTP}/1.0},
	url = {https://datatracker.ietf.org/doc/rfc1945},
	doi = {10.17487/RFC1945},
	urldate = {2026-04-23},
	publisher = {Internet Engineering Task Force},
	author = {Berners-Lee, Tim and Fielding, Roy T. and Nielsen, Henrik},
	month = may,
	year = {1996},
	note = {RFC 1945, https://datatracker.ietf.org/doc/rfc1945},
}

@misc{dierks_tls_1999,
	type = {Request for {Comments}},
	title = {The {TLS} {Protocol} {Version} 1.0},
	url = {https://datatracker.ietf.org/doc/rfc2246},
	doi = {10.17487/RFC2246},
	urldate = {2026-04-23},
	publisher = {Internet Engineering Task Force},
	author = {Dierks, Tim and Allen, Christopher},
	month = jan,
	year = {1999},
	note = {RFC 2246, https://datatracker.ietf.org/doc/rfc2246},
}

@misc{rescorla_http_2000,
	type = {Request for {Comments}},
	title = {{HTTP} {Over} {TLS}},
	url = {https://datatracker.ietf.org/doc/rfc2818},
	doi = {10.17487/RFC2818},
	urldate = {2026-04-23},
	publisher = {Internet Engineering Task Force},
	author = {Rescorla, Eric},
	month = may,
	year = {2000},
	note = {RFC 2818, https://datatracker.ietf.org/doc/rfc2818},
}

@misc{deering_internet_1998,
	type = {Request for {Comments}},
	title = {Internet {Protocol}, {Version} 6 ({IPv6}) {Specification}},
	url = {https://datatracker.ietf.org/doc/rfc2460},
	doi = {10.17487/RFC2460},
	urldate = {2026-04-23},
	publisher = {Internet Engineering Task Force},
	author = {Deering, Steve E. and Hinden, Bob},
	month = dec,
	year = {1998},
	note = {RFC 2460, https://datatracker.ietf.org/doc/rfc2460},
}

@misc{arends_dns_2005,
	type = {Request for {Comments}},
	title = {{DNS} {Security} {Introduction} and {Requirements}},
	url = {https://datatracker.ietf.org/doc/rfc4033},
	doi = {10.17487/RFC4033},
	urldate = {2026-04-23},
	publisher = {Internet Engineering Task Force},
	author = {Arends, Roy and Austein, Rob and Larson, Matt and Massey, Dan and Rose, Scott},
	month = mar,
	year = {2005},
	note = {RFC 4033, https://datatracker.ietf.org/doc/rfc4033},
}

@misc{arends_resource_2005,
	type = {Request for {Comments}},
	title = {Resource {Records} for the {DNS} {Security} {Extensions}},
	url = {https://datatracker.ietf.org/doc/rfc4034},
	doi = {10.17487/RFC4034},
	urldate = {2026-04-23},
	publisher = {Internet Engineering Task Force},
	author = {Arends, Roy and Austein, Rob and Larson, Matt and Massey, Dan and Rose, Scott},
	month = mar,
	year = {2005},
	note = {RFC 4034, https://datatracker.ietf.org/doc/rfc4034},
}

@misc{arends_protocol_2005,
	type = {Request for {Comments}},
	title = {Protocol {Modifications} for the {DNS} {Security} {Extensions}},
	url = {https://datatracker.ietf.org/doc/rfc4035},
	doi = {10.17487/RFC4035},
	urldate = {2026-04-23},
	publisher = {Internet Engineering Task Force},
	author = {Arends, Roy and Austein, Rob and Larson, Matt and Massey, Dan and Rose, Scott},
	month = mar,
	year = {2005},
	note = {RFC 4035, https://datatracker.ietf.org/doc/rfc4035},
}

@misc{lepinski_infrastructure_2012,
	type = {Request for {Comments}},
	title = {An {Infrastructure} to {Support} {Secure} {Internet} {Routing}},
	url = {https://datatracker.ietf.org/doc/rfc6480},
	doi = {10.17487/RFC6480},
	urldate = {2026-04-23},
	publisher = {Internet Engineering Task Force},
	author = {Lepinski, Matt and Kent, Stephen},
	month = feb,
	year = {2012},
	note = {RFC 6480, https://datatracker.ietf.org/doc/rfc6480},
}

@misc{cerf_specification_1974,
	type = {Request for {Comments}},
	title = {Specification of {Internet} {Transmission} {Control} {Program}},
	url = {https://datatracker.ietf.org/doc/rfc675},
	doi = {10.17487/RFC0675},
	urldate = {2026-04-23},
	publisher = {Internet Engineering Task Force},
	author = {Cerf, Vinton G. and Dalal, Yogen and Sunshine, Carl},
	month = dec,
	year = {1974},
	note = {RFC 675, https://datatracker.ietf.org/doc/rfc675},
}

@article{samuelson_status_1988,
	title = {Status quo bias in decision making},
	volume = {1},
	issn = {1573-0476},
	url = {https://doi.org/10.1007/BF00055564},
	doi = {10.1007/BF00055564},
	language = {en},
	number = {1},
	urldate = {2026-04-27},
	journal = {Journal of Risk and Uncertainty},
	author = {Samuelson, William and Zeckhauser, Richard},
	month = mar,
	year = {1988},
	pages = {7--59},
}

@inproceedings{redmiles_dancing_2018,
	address = {New York, NY, USA},
	series = {{EC} '18},
	title = {Dancing {Pigs} or {Externalities}? {Measuring} the {Rationality} of {Security} {Decisions}},
	isbn = {978-1-4503-5829-3},
	shorttitle = {Dancing {Pigs} or {Externalities}?},
	url = {https://dl.acm.org/doi/10.1145/3219166.3219185},
	doi = {10.1145/3219166.3219185},
	urldate = {2026-04-27},
	booktitle = {Proceedings of the 2018 {ACM} {Conference} on {Economics} and {Computation}},
	publisher = {Association for Computing Machinery},
	author = {Redmiles, Elissa M. and Mazurek, Michelle L. and Dickerson, John P.},
	month = jun,
	year = {2018},
	pages = {215--232},
}

@article{kim_investigating_2009,
	title = {Investigating user resistance to information systems implementation: a status quo bias perspective},
	volume = {33},
	issn = {0276-7783},
	shorttitle = {Investigating user resistance to information systems implementation},
	number = {3},
	journal = {MIS Q.},
	author = {Kim, Hee-Woong and Kankanhalli, Atreyi},
	month = sep,
	year = {2009},
	pages = {567--582},
}

@article{polites_shackled_2012,
	title = {Shackled to the {Status} {Quo}: {The} {Inhibiting} {Effects} of {Incumbent} {System} {Habit}, {Switching} {Costs}, and {Inertia} on {New} {System} {Acceptance1}},
	volume = {36},
	issn = {0276-7783},
	shorttitle = {Shackled to the {Status} {Quo}},
	url = {https://dx.doi.org/10.2307/41410404},
	doi = {10.2307/41410404},
	language = {en},
	number = {1},
	urldate = {2026-04-27},
	journal = {Management Information Systems Quarterly},
	publisher = {MIS Quarterly},
	author = {Polites, Greta L. and Karahanna, Elena},
	month = mar,
	year = {2012},
	pages = {21--42},
}

@techreport{barker_considerations_2025,
	address = {Gaithersburg, MD},
	title = {Considerations for {Achieving} {Crypto} {Agility}: {Strategies} and {Practices}:},
	shorttitle = {Considerations for {Achieving} {Crypto} {Agility}},
	url = {https://csrc.nist.gov/pubs/cswp/39/considerations-for-achieving-cryptographic-agility/final},
	doi = {10.6028/NIST.CSWP.39},
	language = {en},
	number = {CSWP 39},
	urldate = {2026-04-27},
	institution = {National Institute of Standards and Technology},
	author = {Barker, Elaine and Chen, Lily and Cooper, David and Moody, Dustin and Regenscheid, Andrew and Souppaya, Murugiah and Newhouse, Bill and Housley, Russ and Turner, Sean and Barker, William and Kent, Karen},
	month = dec,
	year = {2025},
	pages = {39},
}

@inproceedings{jager_one_2013,
	title = {One {Bad} {Apple}: {Backwards} {Compatibility} {Attacks} on {State}-of-the-{Art} {Cryptography}},
	url = {https://www.ndss-symposium.org/ndss2013/ndss-2013-programme/one-bad-apple-backwards-compatibility-attacks-state-art-cryptography/},
	language = {en},
	author = {Jager, Tibor and Paterson, Kenneth G and Holloway, Royal and Somorovsky, Juraj},
	month = apr,
	year = {2013},
}

@misc{parecki_lets_2025,
	title = {Let's fix {OAuth} in {MCP}},
	url = {https://aaronparecki.com/2025/04/03/15/oauth-for-model-context-protocol},
	urldate = {2026-04-27},
	author = {Parecki, Aaron},
	month = apr,
	year = {2025},
	note = {https://aaronparecki.com/2025/04/03/15/oauth-for-model-context-protocol},
}

@misc{delimarsky_whats_2025,
	title = {What's {New} {In} {The} 2025-11-25 {MCP} {Authorization} {Spec}},
	url = {https://den.dev/blog/mcp-november-authorization-spec/},
	language = {en},
	urldate = {2026-04-27},
	author = {Delimarsky, Den},
	month = nov,
	year = {2025},
	note = {https://den.dev/blog/mcp-november-authorization-spec/},
}

@misc{richer_oauth_2015,
	type = {Request for {Comments}},
	title = {{OAuth} 2.0 {Dynamic} {Client} {Registration} {Protocol}},
	url = {https://datatracker.ietf.org/doc/rfc7591},
	doi = {10.17487/RFC7591},
	urldate = {2026-04-27},
	publisher = {Internet Engineering Task Force},
	author = {Jones, Michael B. and Bradley, John and Machulak, Maciej and Hunt, Phil},
	editor = {Richer, Justin},
	month = jul,
	year = {2015},
	note = {RFC 7591, https://datatracker.ietf.org/doc/rfc7591},
}

@misc{soria_parra_governance_2025,
	title = {Governance {Succession} and {Amendment} {Procedures}},
	shorttitle = {{SEP}-2085},
	url = {https://modelcontextprotocol.io/seps/2085-governance-succession-and-amendment},
	language = {en},
	urldate = {2026-04-28},
	author = {Soria Parra, David},
	month = dec,
	year = {2025},
	note = {SEP-2085, https://modelcontextprotocol.io/seps/2085-governance-succession-and-amendment},
}

@article{russell_internet_2013,
	title = {The internet that wasn't},
	volume = {50},
	issn = {1939-9340},
	url = {https://ieeexplore.ieee.org/document/6565559},
	doi = {10.1109/MSPEC.2013.6565559},
	number = {8},
	urldate = {2026-04-28},
	journal = {IEEE Spectrum},
	author = {Russell, Andrew L.},
	month = aug,
	year = {2013},
	pages = {39--43},
}

@article{russell_rough_2006,
	title = {'{Rough} {Consensus} and {Running} {Code}' and the {Internet}-{OSI} {Standards} {War}},
	volume = {28},
	issn = {1934-1547},
	url = {https://ieeexplore.ieee.org/document/1677461},
	doi = {10.1109/MAHC.2006.42},
	number = {3},
	urldate = {2026-04-28},
	journal = {IEEE Annals of the History of Computing},
	author = {Russell, A.L.},
	month = jul,
	year = {2006},
	pages = {48--61},
}

@book{wooldridge_introduction_2009,
	edition = {Second},
	title = {An {Introduction} to {MultiAgent} {Systems}},
	isbn = {978-0-470-51946-2},
	url = {https://www.cs.ox.ac.uk/people/michael.wooldridge/pubs/imas//IMAS2e.html},
	urldate = {2026-04-29},
	publisher = {John Wiley \& Sons},
	author = {Wooldridge, Michael},
	month = may,
	year = {2009},
}

@techreport{pelkey_interview_1988,
	title = {Interview of {Steve} {Crocker}},
	url = {https://archive.computerhistory.org/resources/access/text/2017/12/102738690-05-01-acc.pdf},
	urldate = {2026-04-29},
	institution = {Computer History Museum},
	author = {Pelkey, James L.},
	month = apr,
	year = {1988},
	note = {https://archive.computerhistory.org/resources/access/text/2017/12/102738690-05-01-acc.pdf},
}

@misc{clark_cloudy_1992,
	address = {M.I.T Laboratory for Computer Science},
	title = {A {Cloudy} {Crystal} {Ball} -- {Visions} of the {Future}},
	url = {https://groups.csail.mit.edu/ana/People/DDC/future_ietf_92.pdf},
	language = {en},
	author = {Clark, David D},
	month = jul,
	year = {1992},
	note = {https://groups.csail.mit.edu/ana/People/DDC/future\_ietf\_92.pdf},
}

@misc{nannini_ai_2026,
	title = {{AI} {Agents} {Under} {EU} {Law}},
	url = {https://arxiv.org/abs/2604.04604v1},
	doi = {10.48550/arXiv.2604.04604},
	language = {en},
	urldate = {2026-04-29},
	publisher = {arXiv},
	author = {Nannini, Luca and Smith, Adam Leon and Maggini, Michele Joshua and Panai, Enrico and Feliciano, Sandra and Tiulkanov, Aleksandr and Maran, Elena and Gealy, James and Bisconti, Piercosma},
	month = apr,
	year = {2026},
	note = {arXiv:2604.04604v1 [cs]},
}

@article{nisha_how_2025,
	title = {How {Linux} {Foundation} {Unifies} {Open} {Standards} for {AI} {Agents}},
	url = {https://aimagazine.com/news/one-rule-for-all-agents-linux-foundation-launches-aaif},
	language = {en},
	urldate = {2026-04-29},
	journal = {AI Magazine},
	author = {Nisha, Rithula},
	month = dec,
	year = {2025},
}

@misc{waitzman_standard_1990,
	type = {Request for {Comments}},
	title = {Standard for the transmission of {IP} datagrams on avian carriers},
	url = {https://datatracker.ietf.org/doc/rfc1149},
	doi = {10.17487/RFC1149},
	urldate = {2026-04-30},
	publisher = {Internet Engineering Task Force},
	author = {Waitzman, David},
	month = apr,
	year = {1990},
	note = {RFC 1149, https://datatracker.ietf.org/doc/rfc1149},
}

@misc{hardt_oauth_2012,
	type = {Request for {Comments}},
	title = {The {OAuth} 2.0 {Authorization} {Framework}},
	url = {https://datatracker.ietf.org/doc/rfc6749},
	doi = {10.17487/RFC6749},
	urldate = {2026-04-30},
	publisher = {Internet Engineering Task Force},
	author = {Hardt, Dick},
	month = oct,
	year = {2012},
	note = {RFC 6749, https://datatracker.ietf.org/doc/rfc6749},
}

@misc{soria_parra_mcp_2026,
	title = {{MCP} {Contributor} {Ladder}},
	shorttitle = {{SEP}-2148},
	url = {https://modelcontextprotocol.io/seps/2148-contributor-ladder},
	language = {en},
	urldate = {2026-04-30},
	author = {Soria Parra, David and Novotny, Sarah},
	month = jan,
	year = {2026},
	note = {SEP-2148, https://modelcontextprotocol.io/seps/2148-contributor-ladder},
}

@misc{nalley_aaifs_2026,
	title = {{AAIF}’s {First} {Quarter} {Success} {Story}: {New} {Members}, {Technical} {Wins}, and {Open} {Governance}},
	shorttitle = {{AAIF}’s {First} {Quarter} {Success} {Story}},
	url = {https://aaif.io/blog/aaifs-first-quarter-success-story-new-members-technical-wins-and-open-governance/},
	language = {en-US},
	urldate = {2026-05-01},
	journal = {Agentic AI Foundation (AAIF)},
	author = {Nalley, David},
	month = feb,
	year = {2026},
	note = {https://aaif.io/blog/aaifs-first-quarter-success-story-new-members-technical-wins-and-open-governance/},
}

@misc{zakon_hobbes_1997,
	type = {Request for {Comments}},
	title = {Hobbes' {Internet} {Timeline}},
	url = {https://datatracker.ietf.org/doc/rfc2235},
	doi = {10.17487/RFC2235},
	urldate = {2026-05-01},
	publisher = {Internet Engineering Task Force},
	author = {Zakon, Robert H.},
	month = nov,
	year = {1997},
	note = {RFC 2235, https://datatracker.ietf.org/doc/rfc2235},
}

@misc{biddle_no_2016,
	address = {Rochester, NY},
	type = {{SSRN} {Scholarly} {Paper}},
	title = {No {Standard} for {Standards}: {Understanding} the {ICT} {Standards} {Development} {Ecosystem}},
	shorttitle = {No {Standard} for {Standards}},
	url = {https://papers.ssrn.com/abstract=3023650},
	doi = {10.2139/ssrn.3023650},
	language = {en},
	urldate = {2026-05-20},
	publisher = {Social Science Research Network},
	author = {Biddle, Brad},
	month = jun,
	year = {2016},
}

@misc{itu_requirements_2025,
	title = {Requirements and evaluation methods of artificial intelligence agents based on large scale pre-trained models},
	url = {https://www.itu.int/rec/T-REC-F.748.46-202503-I/en},
	urldate = {2026-05-20},
	author = {ITU},
	month = mar,
	year = {2025},
}

@misc{van_draanen_glismann_lexicon_2025,
	title = {Lexicon: {How} {China} talks about 'agentic {AI}'},
	url = {https://digichina.stanford.edu/work/lexicon-how-china-talks-about-agentic-ai/},
	language = {en},
	urldate = {2026-05-20},
	journal = {DigiChina},
	author = {van Draanen Glismann, Malou and Webster, Graham},
	month = oct,
	year = {2025},
	note = {https://digichina.stanford.edu/work/lexicon-how-china-talks-about-agentic-ai/},
}

@misc{crocker_official_1970,
	type = {Request for {Comments}},
	title = {Official {Protocol} {Proffering}},
	shorttitle = {{RFC} 54},
	url = {https://datatracker.ietf.org/doc/html/rfc54},
	language = {en},
	urldate = {2026-05-20},
	author = {Crocker, Steve and Postel, Dr Jon and Newkirk, J. and Kraley, M.},
	month = jun,
	year = {1970},
	note = {RFC 54, https://datatracker.ietf.org/doc/rfc54},
}

@misc{postel_internet_1981,
	type = {Request for {Comments}},
	title = {Internet {Protocol}},
	shorttitle = {{RFC} 791},
	url = {https://datatracker.ietf.org/doc/html/rfc791},
	language = {en},
	urldate = {2026-05-20},
	author = {Postel, Jon},
	month = sep,
	year = {1981},
	note = {RFC 791, https://datatracker.ietf.org/doc/rfc791},
}

@misc{postel_transmission_1981,
	type = {Request for {Comments}},
	title = {Transmission {Control} {Protocol}},
	shorttitle = {{RFC} 793},
	url = {https://datatracker.ietf.org/doc/html/rfc793},
	language = {en},
	urldate = {2026-05-20},
	author = {Postel, Jon},
	month = sep,
	year = {1981},
	note = {RFC 793, https://datatracker.ietf.org/doc/rfc793},
}

@misc{cerf_internet_2002,
	type = {Request for {Comments}},
	title = {The {Internet} is for {Everyone}},
	url = {https://datatracker.ietf.org/doc/rfc3271},
	doi = {10.17487/RFC3271},
	urldate = {2026-05-20},
	publisher = {Internet Engineering Task Force},
	author = {Cerf, Vinton G.},
	month = apr,
	year = {2002},
	note = {RFC 3271,},
}

@misc{carpenter_architectural_1996,
	type = {Request for {Comments}},
	title = {Architectural {Principles} of the {Internet}},
	url = {https://datatracker.ietf.org/doc/rfc1958},
	doi = {10.17487/RFC1958},
	urldate = {2026-05-20},
	publisher = {Internet Engineering Task Force},
	author = {Carpenter, Brian E.},
	month = jun,
	year = {1996},
	note = {RFC 1958, https://datatracker.ietf.org/doc/rfc1958},
}

@misc{iab_ethics_1989,
	type = {Request for {Comments}},
	title = {Ethics and the {Internet}},
	url = {https://datatracker.ietf.org/doc/rfc1087},
	doi = {10.17487/RFC1087},
	urldate = {2026-05-20},
	publisher = {Internet Engineering Task Force},
	author = {IAB},
	month = jan,
	year = {1989},
	note = {RFC 1087,},
}

@misc{postel_instructions_1997,
	type = {Request for {Comments}},
	title = {Instructions to {RFC} {Authors}},
	url = {https://datatracker.ietf.org/doc/rfc2223},
	doi = {10.17487/RFC2223},
	urldate = {2026-05-20},
	publisher = {Internet Engineering Task Force},
	author = {Postel, Jon and Reynolds, Joyce K.},
	month = oct,
	year = {1997},
	note = {RFC 2223, https://datatracker.ietf.org/doc/rfc2223},
}

@article{schmidt_democracy_2013,
	title = {Democracy and {Legitimacy} in the {European} {Union} {Revisited}: {Input}, {Output} and ‘{Throughput}’},
	volume = {61},
	copyright = {© 2012 The Author. Political Studies © 2012 Political Studies Association},
	issn = {1467-9248},
	shorttitle = {Democracy and {Legitimacy} in the {European} {Union} {Revisited}},
	url = {https://onlinelibrary.wiley.com/doi/abs/10.1111/j.1467-9248.2012.00962.x},
	doi = {10.1111/j.1467-9248.2012.00962.x},
	language = {en},
	number = {1},
	urldate = {2026-08-11},
	journal = {Political Studies},
	author = {Schmidt, Vivien A.},
	year = {2013},
	pages = {2--22},
}

@book{palladino_legitimacy_2021,
	address = {Cham},
	title = {Legitimacy, {Power}, and {Inequalities} in the {Multistakeholder} {Internet} {Governance}: {Analyzing} {IANA} {Transition}},
	copyright = {http://www.springer.com/tdm},
	isbn = {978-3-030-56130-7 978-3-030-56131-4},
	shorttitle = {Legitimacy, {Power}, and {Inequalities} in the {Multistakeholder} {Internet} {Governance}},
	url = {http://link.springer.com/10.1007/978-3-030-56131-4},
	doi = {10.1007/978-3-030-56131-4},
	language = {en},
	urldate = {2026-08-11},
	publisher = {Springer International Publishing},
	author = {Palladino, Nicola and Santaniello, Mauro},
	year = {2021},
}

@book{scharpf_governing_1999,
	title = {Governing in {Europe}: {Effective} and {Democratic}?},
	isbn = {978-0-19-829545-7},
	shorttitle = {Governing in {Europe}},
	url = {https://doi.org/10.1093/acprof:oso/9780198295457.001.0001},
	doi = {10.1093/acprof:oso/9780198295457.001.0001},
	urldate = {2026-08-11},
	publisher = {Oxford University Press},
	author = {Scharpf, Fritz},
	month = mar,
	year = {1999},
}

@misc{zhou_first_2026,
	title = {A {First} {Measurement} {Study} on {Authentication} {Security} in {Real}-{World} {Remote} {MCP} {Servers}},
	url = {http://arxiv.org/abs/2605.22333},
	doi = {10.48550/arXiv.2605.22333},
	urldate = {2026-09-08},
	publisher = {arXiv},
	author = {Zhou, Huijun and Zhang, Xiaohan and Zhang, Haozhe and Zhang, Haoyang and Zhang, Mi and Yang, Min},
	month = may,
	year = {2026},
	note = {arXiv:2605.22333 [cs.CR]},
}

@misc{ox_security_mother_2026,
	title = {The {Mother} of {All} {AI} {Supply} {Chains}: {Anthropic}’s "{By} {Design}" failure at the heart of the {AI} ecosystem},
	shorttitle = {The {Mother} of {All} {AI} {Supply} {Chains}},
	url = {https://www.ox.security/reports/the-mother-of-all-ai-supply-chains-anthropics-by-design-failure-at-the-heart-of-the-ai-ecosystem/},
	language = {en-US},
	urldate = {2026-09-08},
	journal = {OX Security},
	author = {OX Security},
	month = apr,
	year = {2026},
}

@techreport{faaborg-anderson_geopolitics_2022,
	address = {Cambridge, MA, USA},
	title = {The {Geopolitics} of {Digital} {Standards}},
	url = {https://www.belfercenter.org/publication/geopolitics-digital-standards},
	language = {en},
	urldate = {2026-09-08},
	institution = {The Belfer Center for Science and International Affairs},
	author = {Faaborg-Anderson, Sophie and Temes, Lindsay},
	month = jul,
	year = {2022},
}

@inproceedings{khare_web_2022,
	title = {The web we weave: {Untangling} the social graph of the {IETF}},
	volume = {16},
	shorttitle = {The web we weave},
	url = {https://ojs.aaai.org/index.php/ICWSM/article/view/19310},
	urldate = {2026-10-05},
	booktitle = {Proceedings of the {International} {AAAI} {Conference} on {Web} and {Social} {Media}},
	author = {Khare, Prashant and Karan, Mladen and McQuistin, Stephen and Perkins, Colin and Tyson, Gareth and Purver, Matthew and Healey, Patrick and Castro, Ignacio},
	year = {2022},
	pages = {500--511},
}

@incollection{cath_send_2026,
	series = {Digital {Humanities} {Series}},
	title = {Send {Conservatively}, {Accept} {Liberally}: {Cultural} {Governance} of {Open} {Internet} {Standards}},
	volume = {6},
	issn = {2054-2410},
	shorttitle = {1. {Send} {Conservatively}, {Accept} {Liberally}},
	url = {https://www.openbookpublishers.com/books/10.11647/obp.0528/chapters/10.11647/obp.0528.01},
	doi = {10.11647/obp.0528.01},
	language = {English},
	urldate = {2026-10-05},
	booktitle = {The {Politics} of {Open} {Infrastructures}: {Power}, {Governance}, and {Justice} in {Digital} {Knowledge} {Practices}},
	publisher = {Open Book Publishers},
	author = {Cath, Corinne},
	collaborator = {Mayer, Katja and Mager, Astrid and Ridgway, Renée},
	month = may,
	year = {2026},
	pages = {21--42},
}

@misc{kumar_robustness_2025,
	title = {Robustness in {Large} {Language} {Models}: {A} {Survey} of {Mitigation} {Strategies} and {Evaluation} {Metrics}},
	shorttitle = {Robustness in {Large} {Language} {Models}},
	url = {http://arxiv.org/abs/2505.18658},
	doi = {10.48550/arXiv.2505.18658},
	urldate = {2026-10-06},
	publisher = {arXiv},
	author = {Kumar, Pankaj and Mishra, Subhankar},
	month = nov,
	year = {2025},
	note = {arXiv:2505.18658 [cs.CL]},
}

\end{document}